\pdfoutput=1
\documentclass[aps,prb,reprint,longbibliography,superscriptaddress,floatfix]{revtex4-2}
\usepackage[T1]{fontenc}
\usepackage[utf8]{inputenc}
\usepackage{amsmath,amssymb,graphicx,booktabs,microtype}
\usepackage[colorlinks=true,linkcolor=blue,citecolor=blue,urlcolor=blue]{hyperref}
\hypersetup{pdftitle={Reference-state dependence and interface scaling of superfluid response in a two-dimensional crystal}}
\hypersetup{pdfauthor={N\'estor Sep\'ulveda}}
\newcommand{\av}[1]{\left\langle #1\right\rangle}
\newcommand{\dd}{\mathrm{d}}
\newcommand{\fs}{\boldsymbol{f}^{\mathrm{s}}}
\newcommand{\fbar}{\overline{f}^{\mathrm{s}}}
\newcommand{\LowAngleToleranceChange}{2.653\times10^{-8}}
\newcommand{\LowAngleResidual}{5.926\times10^{-11}}
\newcommand{\LowAngleGridChange}{2.438\times10^{-7}}

\newcommand{\SparseControlMean}{0.00777529}
\newcommand{\SparseDefectMean}{0.00814854}
\newcommand{\SparseEnhancement}{4.80}

\begin{document}
\title{Reference-state dependence and interface scaling of superfluid response in a two-dimensional crystal}
\author{N\'estor Sep\'ulveda}
\email[Contact author: ]{nesepulv@gmail.com}
\affiliation{School of Engineering and Sciences, Universidad Adolfo Ib\'a\~nez, Diagonal las Torres 2640, Pe\~nalol\'en, Santiago, Chile}

\begin{abstract}
Grain boundaries modify both transport paths and the surrounding crystal, making their contribution to superfluid stiffness difficult to isolate. We quantify these contributions in stationary two-dimensional soft-core Gross--Pitaevskii (GP) crystals using tensor response, independently prepared bicrystals, and measured interface lengths. A one-percent change of lattice spacing alters the bulk stiffness by approximately 5--9\%, comparable to the boundary contribution. For the same defective state, two bulk references give relative excesses of 0.027\% and 4.15\%. Matching the comparator to grain strain and chemical potential yields positive diagonal excesses in paired preparations at three interaction strengths. Within a twelve-state family with comparable grain interiors, the excess is robust to 288 bulk substitutions and follows an approximate length-per-area relation. Independent fixed-area pairs yield a line-response coefficient 3.1\% higher than the three-size estimate. Size and directional dependence restrict this relation to the specified crystal conditions. Reduced link networks relate the directional response to the organization of boundary connections but do not accurately resolve every small bulk-subtracted excess. The results distinguish boundary enhancement from changes in the surrounding crystal and establish a quantitative comparison protocol for superfluid transport in defective crystals.
\end{abstract}
\maketitle
\section{Introduction}
Grain boundaries couple crystalline deformation to transport through a spatially inhomogeneous superfluid. The resulting response depends on the interface and on the state of the surrounding crystal. Separating these contributions is necessary to assign a superfluid stiffness to a defect. Density-based constraints on superfluidity were established by Leggett~\cite{Leggett1970,Leggett1998}, and microscopic mechanisms are reviewed in Ref.~\cite{Boninsegni2012}. Measurements in spatially modulated condensates~\cite{Tao2023,Chauveau2023} and dipolar supersolids~\cite{Biagioni2024,Norcia2021} now give experimental access to the superfluid fraction of modulated states.

The nonlocal Gross--Pitaevskii (GP) model used here originates in Pomeau and Rica~\cite{Pomeau1994}. Its elastic and phase-response formulation was developed by Josserand, Pomeau and Rica~\cite{Josserand2007}. A one-dimensional analysis was presented in Ref.~\cite{Sepulveda2008}. Blakie~\cite{Blakie2024} computed the superfluid tensor of triangular, square and stripe states in the same two-dimensional setting, and Rakic, Ho and Lee~\cite{Rakic2024} obtained elastic moduli and superfluid response together by homogenization. We use these formulations for the response of homogeneous deformed crystals.

Grain-boundary enhancement of the superfluid fraction was reported in Ref.~\cite{Sepulveda2010}. That work found increases up to 50\% for defect networks and related the excess to boundary length per area, $\Delta f^{\mathrm s}\sim a/L$. Grain-boundary superflow has also been found by quantum Monte Carlo in helium~\cite{Pollet2007} and tilted lattice dipoles~\cite{Zhang2022}. Occupation-defect supersolidity in soft-core bosons~\cite{Cinti2014} is a distinct many-body mechanism.

An excess is defined relative to a bulk state. In interface thermodynamics, Gibbs~\cite{Gibbs1878} formulated surface excesses relative to extrapolated bulk phases and a dividing surface, and Cahn~\cite{Cahn1979} generalized the construction to avoid an arbitrary dividing surface and to include solid-state deformation. Superfluid stiffness is not a Gibbs surface excess, because the phase relaxes throughout the sample, but the same question arises: which bulk crystal is subtracted?

Here we quantify how much the inferred grain-boundary excess changes with its comparator, relate this sensitivity to microscopic transport, and use controlled comparators to test the length-per-area scaling of Ref.~\cite{Sepulveda2010}. Independent preparations, measured interface lengths and bulk references matched to deformation and density make the comparison explicit. Controlled lattice deformations quantify the bulk sensitivity, and a reduced link network connects the spatial organization of inter-peak transport to the directional boundary response.

\section{Model and definitions}
\subsection{Nonlocal interaction and units}
We consider a complex field $\psi$ on a square periodic domain $\Omega=[0,L)^2$, with spatial average $\av{g}=L^{-2}\int_\Omega g\,\dd^2x$. In units $\hbar=m=1$, with $\hbar$ the reduced Planck constant and $m$ the particle mass, the energy density is
\begin{equation}
 \mathcal E[\psi]=\frac12\av{|\nabla\psi|^2}
 +\frac12\av{\rho\,(U*\rho)},\qquad \rho=|\psi|^2,
 \label{eq:energy}
\end{equation}
where $U(r)=U_0\Theta(a-r)$, with $\Theta$ the Heaviside step function, is periodized on the domain~\cite{Pomeau1994,Josserand2007}. The coupling is
\begin{equation}
 \Lambda=n_0\pi a^4U_0,\qquad
 \widehat U(k)=\begin{cases}2\pi U_0a J_1(ka)/k,&k>0,\\ \pi a^2U_0,&k=0,\end{cases}
 \label{eq:kernel}
\end{equation}
Here $(U*\rho)(\boldsymbol x)=\int_\Omega U_{\rm per}(\boldsymbol x-\boldsymbol x')\rho(\boldsymbol x')\,\dd^2x'$, where $U_{\rm per}$ is the periodized kernel. Its Fourier multiplier is the unnormalized plane transform $\widehat U(\boldsymbol k)=\int_{\mathbb R^2}U(|\boldsymbol x|)e^{-i\boldsymbol k\cdot\boldsymbol x}\,\dd^2x$, evaluated at $\boldsymbol k=2\pi\boldsymbol n/L$, $\boldsymbol n\in\mathbb Z^2$; $k=|\boldsymbol k|$. The mean density is $n_0=\av{\rho}$, $a$ is the interaction range, $U_0$ its amplitude, and $J_1$ the order-one Bessel function of the first kind. The primary fixed-density ensemble uses $a=8$, $n_0=1$ and $\Lambda=128.68$, hence $U_0=0.01000002836$. The interior-controlled ensemble of Sec.~\ref{sec:controlled} keeps this $U_0$ and varies the mean density, with $\Lambda_0=\pi a^4U_0$ and $\Lambda=n_0\Lambda_0$. For a uniform field, the chemical potential is $\mu=\Lambda/a^2$ and $\mathcal E=n_0\Lambda/(2a^2)$.

At $n_0=1$, the scaled position $\boldsymbol\xi=\boldsymbol x/a$ and time $\tau=t/a^2$ give the dynamics
\begin{equation}
 i\partial_\tau\psi=-\tfrac12\nabla_\xi^2\psi+
 \frac{\Lambda}{\pi}\psi\int_{|\boldsymbol\xi-\boldsymbol\xi'|<1}\rho(\boldsymbol\xi')\,\dd^2\xi'.
 \label{eq:scaled}
\end{equation}
The fixed-density ensemble uses $L/a=16,32,48,64$. Box side $L$ and grid size are kept distinct throughout.

\subsection{Stationary crystals}
A positive real zero-current stationary amplitude $u$, normalized by $\av{u^2}=n_0$, satisfies
\begin{equation}
 H[u]u=\mu u,\qquad H[u]=-\tfrac12\nabla^2+V,\quad V=U*u^2.
\end{equation}
Here $\mu$ is the chemical potential enforcing the fixed norm. The square grid has $N_g$ points per direction and spacing $h=L/N_g$; the crystal itself is approximately hexagonal. Local density maxima are called sites or peaks; their number is $N_s$.

\subsection{Global superfluid response}
The response is defined on the branch of stationary solutions of $\mathcal E-\boldsymbol v\cdot\boldsymbol P$, where $\boldsymbol v$ is the boost velocity and $\boldsymbol P$ the momentum density,
\begin{equation}
 P_i=\av{\operatorname{Im}(\psi^*\partial_i\psi)},\qquad
 f^{\mathrm s}_{ij}=\delta_{ij}-\frac{1}{n_0}\left.\frac{\partial P_i}{\partial v_j}\right|_{\boldsymbol v=0},
 \label{eq:boost}
\end{equation}
Here $i,j\in\{x,y\}$, $\partial_i=\partial/\partial x_i$, and $\delta_{ij}$ is the Kronecker delta. Repeated Cartesian indices are summed in tensor contractions; labels $jj$ on directional observables denote a fixed direction without summation. The derivative is taken at fixed mean density, branch and phase-winding number (zero net phase winding across the periodic domain). Writing $\psi_{\boldsymbol v}=u+i\sum_jv_j\eta_j+O(v^2)$ gives
\begin{equation}
 A\eta_j=-\partial_j u,\quad
 A=-\tfrac12\nabla^2+V-\mu,\quad \av{u\eta_j}=0,
 \label{eq:response}
\end{equation}
where the orthogonality constraint removes the phase null mode; density and chemical potential do not vary at linear order. Integration by parts yields
\begin{equation}
 f^{\mathrm s}_{ij}=\delta_{ij}+\frac{2\av{\eta_j\partial_i u}}{n_0}.
 \label{eq:tensor}
\end{equation}
The solve covers the whole defective domain. We report the tensor components, its eigenvalues $f_1\leq f_2$ and $\fbar=\operatorname{tr}\fs/2$, retaining anisotropy as in Refs.~\cite{Saslow1976,Josserand2007,Blakie2024}. The physical stiffness is $K_{ij}=n_0f^{\mathrm s}_{ij}$; it has density units in the chosen units (the phase-energy coefficient in physical units is $\hbar^2K_{ij}/m$). We write $\overline K=\operatorname{tr}K/2$. Equations~\eqref{eq:response}--\eqref{eq:tensor} are equivalent to phase homogenization~\cite{Josserand2007}, with periodic correctors $\theta_j=-\eta_j/u$, which have length units. The phase change under a boost is $-\sum_jv_j\theta_j$; under an imposed phase gradient it is $\sum_jq_j(x_j+\theta_j)$.

\subsection{Local phase stiffness}
The unit-twist flux associated with Eq.~\eqref{eq:response} is
\begin{equation}
 j_i^{(j)}=\rho\delta_{ij}-u\partial_i\eta_j+\eta_j\partial_i u,
 \qquad \av{j_i^{(j)}}=n_0 f^{\mathrm s}_{ij}.
 \label{eq:flux}
\end{equation}
For each fixed direction $j$ (no sum), $\rho=u^2>0$ and $e_j=|\boldsymbol j^{(j)}|^2/\rho$ is a nonnegative local stiffness density, and
\begin{equation}
 \av{e_j}=n_0 f^{\mathrm s}_{jj},
 \label{eq:identity}
\end{equation}
which follows from $\boldsymbol j^{(j)}=\rho(\hat{\boldsymbol x}_j+\nabla\theta_j)$, periodicity and current conservation $\nabla\cdot\boldsymbol j^{(j)}=0$; here $\hat{\boldsymbol x}_j$ is a Cartesian unit vector. Because division by small densities amplifies numerical error, a local map is used only if Eq.~\eqref{eq:identity} holds to a relative accuracy of $10^{-3}$.

\subsection{Excess response with a specified reference}\label{sec:excess-concept}
Let $\mathcal R$ specify the bulk deformation, the density or chemical-potential convention, the grain partition and the response constraints. For a defective state, labeled GB (grain boundary), the assigned excess is
\begin{equation}
\Delta K^{[\mathcal R]}=K^{\rm GB}-K^{\rm bulk,[\mathcal R]}.
\label{eq:reference-excess}
\end{equation}
Changing $\mathcal R$ changes the assigned contribution but not the measured $K^{\rm GB}$. Equivalently, if $\delta\mathcal E_{\rm GB}(\boldsymbol q)$ is the change of stationary energy density relative to its value at $\boldsymbol q=0$, where $\boldsymbol q$ is an imposed phase gradient (inverse length) and the periodic part of the field remains on the zero-winding branch at fixed mean density,
\begin{equation}
\delta\mathcal E_{\rm GB}(\boldsymbol q)
-\tfrac12 q_iK^{\rm bulk,[\mathcal R]}_{ij}q_j
=\tfrac12 q_i\Delta K^{[\mathcal R]}_{ij}q_j+o(|\boldsymbol q|^2),
\label{eq:excess-curvature}
\end{equation}
so the excess is a difference of phase-energy curvatures. The bulk composite is itself obtained from a global phase problem, without assuming local additivity of stiffness. Unlike Cahn's generalized excesses~\cite{Cahn1979}, no invariance under a change of convention is expected. We drop the index $[\mathcal R]$ once the reference is specified.

\section{Preparation ensembles and bulk comparisons}\label{sec:methods}
Stationary states are computed by fixed-norm spectral minimization; the periodic box imposes a commensuration strain on the hexagonal crystal. All states included in the ensemble response estimates satisfy
\begin{equation}
R=\frac{\|(H-\mu)u\|_2}{\max(1,|\mu|)\|u\|_2}\leq10^{-9}.
\label{eq:residual}
\end{equation}
Here $\|g\|_2^2=\int_\Omega|g|^2\,\dd^2x$, and the unit energy in the denominator is expressed in the same numerical units as $\mu$. Grid refinement, finite-boost comparisons and the local-flux identity test the numerical accuracy, and the solver reproduces the triangular-crystal responses reported by Blakie~\cite{Blakie2024} at $\Lambda=40$ and 100 (Appendix~\ref{app:numerics}).

The fixed-density ensemble contains eight independent bicrystals at each $L=128,256,384,512$, with $a=8$, $\Lambda=128.68$ and $h=0.5$. Two nominally $30^\circ$ grain boundaries cross each periodic domain, and random grain translations and boundary perturbations define independent preparations. The interface length $\ell$ is measured on the relaxed state from orientational-order contours (Appendix~\ref{app:preparations}).

We compare prescribed single crystals with homogeneous composites that reproduce the measured grain interiors. For grain $g$, with reference-cell matrix $B_g$ (columns are primitive vectors) and measured deformation gradient $F_g$ mapping that cell to the observed interior, the composite uses
\begin{equation}
K_g=n_gf_g(\Lambda_0 n_g,F_gB_g),
\end{equation}
where $n_g$ is the bulk density assigned to grain $g$ by the chosen reference convention, $\Lambda_0=\pi a^4U_0$, and $f_g(\Lambda,C)$ is the superfluid tensor of the homogeneous primitive cell $C$ at coupling $\Lambda$. All grain tensors are expressed in the box Cartesian frame. On the measured grain partition, set $K_{\rm loc}(\boldsymbol x)=K_g$ in grain $g$. The bulk tensor is defined by
\begin{equation}
\begin{gathered}
\nabla\cdot[K_{\rm loc}(\hat{\boldsymbol x}_j+\nabla w_j)]=0,\\
K^{\rm bulk}_{ij}=\av{[K_{\rm loc}(\hat{\boldsymbol x}_j+\nabla w_j)]_i},
\end{gathered}
\label{eq:bulk-composite}
\end{equation}
with periodic zero-mean correctors $w_j$; the equation holds weakly across grain boundaries. Native-density composites use the measured $n_g$; fixed-total-mass composites constrain the partition-weighted mean to $n_0$, whereas chemical-potential-matched composites determine each $n_g$ from the homogeneous equation of state at the bicrystal $\mu$. They are different bulk states. The excess and interface density are
\begin{equation}
\begin{gathered}
\Delta K_{ij}=K^{\rm GB}_{ij}-K^{\rm bulk}_{ij},\\
\Delta\overline K=\tfrac12\operatorname{tr}\Delta K,\qquad \nu=\ell/L^2.
\end{gathered}
\label{eq:physicalexcess}
\end{equation}
Here $\ell$ is the total length of all retained interfaces, $\nu$ has inverse-length units, and positive $\Delta K_{jj}$ means enhancement in direction $j$. At unit mean density $K$ and $f$ coincide numerically.

We recommend the strain- and chemical-potential-matched composite. It keeps the measured grain deformations and partition, and sets each bulk density from its equation of state at the bicrystal chemical potential. This fixes the mechanical geometry and the particle-exchange condition of the comparator, in the spirit of bulk-state matching in interface thermodynamics~\cite{Gibbs1878,Cahn1979}. Its total mass may differ from that of the bicrystal, so the subtraction uses $K$ rather than fractions normalized by different densities. The prescription is operational; the other references quantify its convention dependence.

Preparation means are denoted $\langle\cdot\rangle_{\rm prep}$, distinct from spatial means; ensemble tables and size-mean plots average preparations at fixed $L$. Preparation spread is measured across independent states, resampling preserves size strata, and numerical continuations are not counted as preparations (Appendix~\ref{app:statistics}).

\section{Grain-boundary response and reference dependence}
\subsection{Spatial response of relaxed bicrystals}
\begin{figure*}[!t]\centering
\includegraphics[width=.76\textwidth]{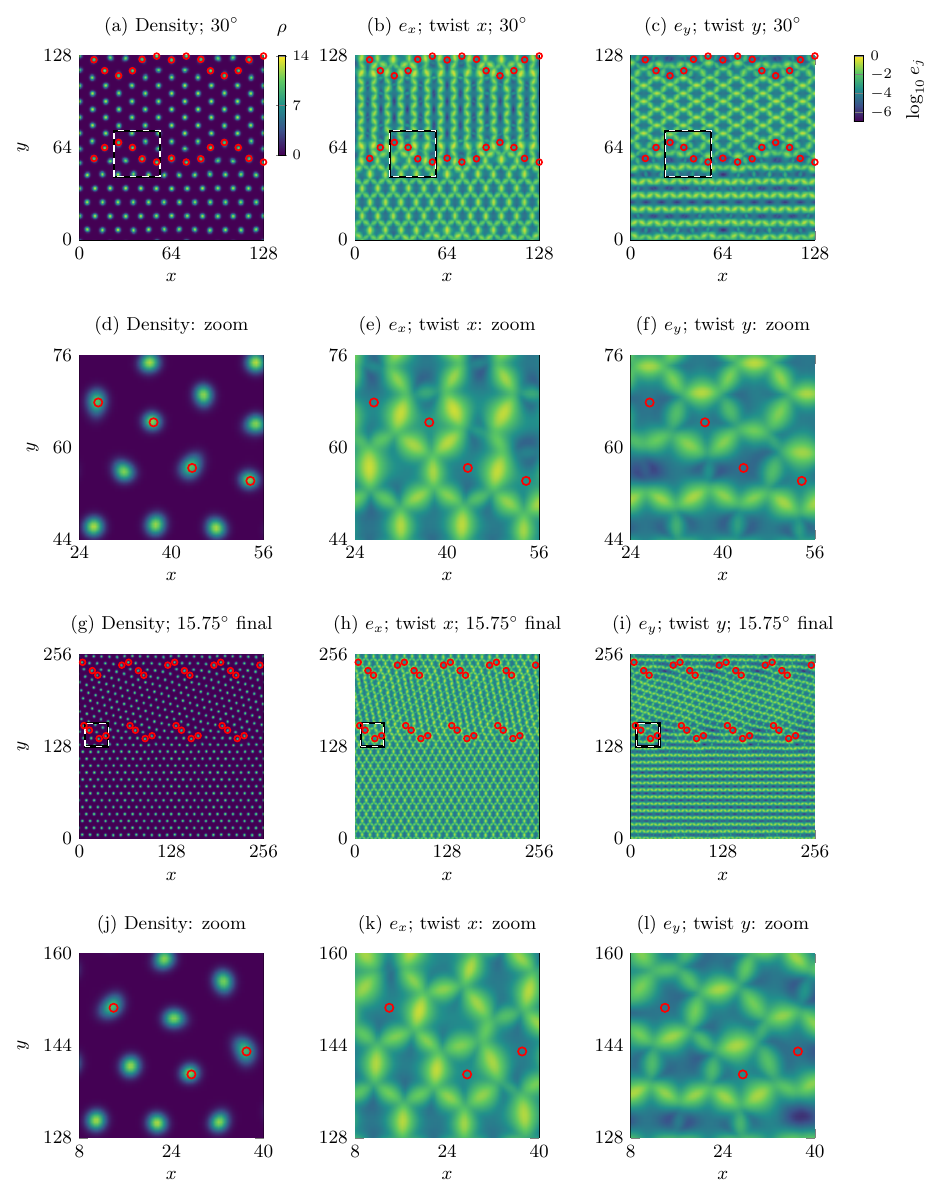}
\caption{Density and local phase stiffness of relaxed bicrystals. Columns show $\rho$ on the linear scale $[0,14]$, $e_x$, and $e_y$; stiffness uses $\log_{10}(e_j/n_0)\in[-7,0]$. Rows 1--2 show the $30^\circ$ family at $L=128$; rows 3--4 show the low-angle family at $L=256$, with final misorientation $15.75^\circ$. Dashed boxes locate the $32\times32$ zoom windows. Red circles mark non-sixfold coordination. Fields are shown without smoothing; stationary residuals are $9.99\times10^{-9}$ and $5.93\times10^{-11}$.}\label{fig:local}
\end{figure*}

Figure~\ref{fig:local} shows density and local stiffness in two relaxed bicrystals. For a mask $\mathcal M$ formed by radius-$a$ disks around sites with non-sixfold coordination, define
\begin{equation}
p_{\mathcal M}=|\mathcal M|/L^2,\qquad
\mathcal Q_j=\frac{\int_{\mathcal M}e_j\,\dd^2x}{p_{\mathcal M}\int_\Omega e_j\,\dd^2x}.
\end{equation}
$|\mathcal M|$ denotes the mask area. All three measured bicrystals have $\mathcal Q_x>1$ and $\mathcal Q_y<1$ (Table~\ref{tab:local}): stiffness for flow along the interfaces is concentrated near them. $\mathcal Q_j$ is a phase-stiffness observable, not a local particle fraction.

\subsection{The same defective state under different references}
A reconstructed low-angle bicrystal at $L=256$ contains 644 density peaks and 32 coordination anomalies. Its response tensor is
\begin{equation}
\fs_{\rm GB}=\begin{pmatrix}0.010264&-0.000079\\-0.000079&0.008682\end{pmatrix},
\quad\fbar_{\rm GB}=0.009473,
\label{eq:measuredtensor}
\end{equation}
with eigenvalue anisotropy $(f_2-f_1)/(f_2+f_1)=0.084$, within the nearly isotropic range of Ref.~\cite{Sepulveda2010}. With the arithmetic comparator and relative excess
\begin{equation}
\begin{gathered}
f_{\rm ref}=\tfrac12(\fbar_A+\fbar_B),\\
\Delta f=\fbar_{\rm GB}-f_{\rm ref},\quad
\mathcal H=100\Delta f/f_{\rm ref},
\end{gathered}
\label{eq:excess}
\end{equation}
where $A$ and $B$ label the two corresponding single-crystal orientations and $\mathcal H$ is the percentage enhancement. Replacing the 672-site reference crystals by 644-site crystals changes $\mathcal H$ from 0.027\% to 4.15\% for the same defective state (Table~\ref{tab:references}). This change is four orders of magnitude larger than the numerical sensitivity. Site-count matching alone still leaves strain and orientation differences; the ensembles below control the grain interiors explicitly.
\begin{table*}[!t]\centering\small
\begin{tabular}{rrrrrr}
\toprule
$N_s^{\rm ref}$ & $\fbar_A$ & $\fbar_B$ & $f_{\rm ref}$ & $\fbar_{\rm GB}$ & $\mathcal H$ (\%)\\
\midrule
672 & 0.00912955 & 0.00981172 & 0.00947063 & 0.00947322 & 0.027\\
644 & 0.00857326 & 0.00961868 & 0.00909597 & 0.00947322 & 4.147\\
\bottomrule\end{tabular}

\caption{Reference sensitivity for the same 644-site bicrystal at $L=256$ and $h=0.5$. $\fbar_A$ and $\fbar_B$ are the single-crystal responses; $f_{\rm ref}$ is their arithmetic mean.}\label{tab:references}
\end{table*}

\subsection{Reference dependence across independent preparations}
\begin{figure*}[!t]\centering\includegraphics[width=\textwidth]{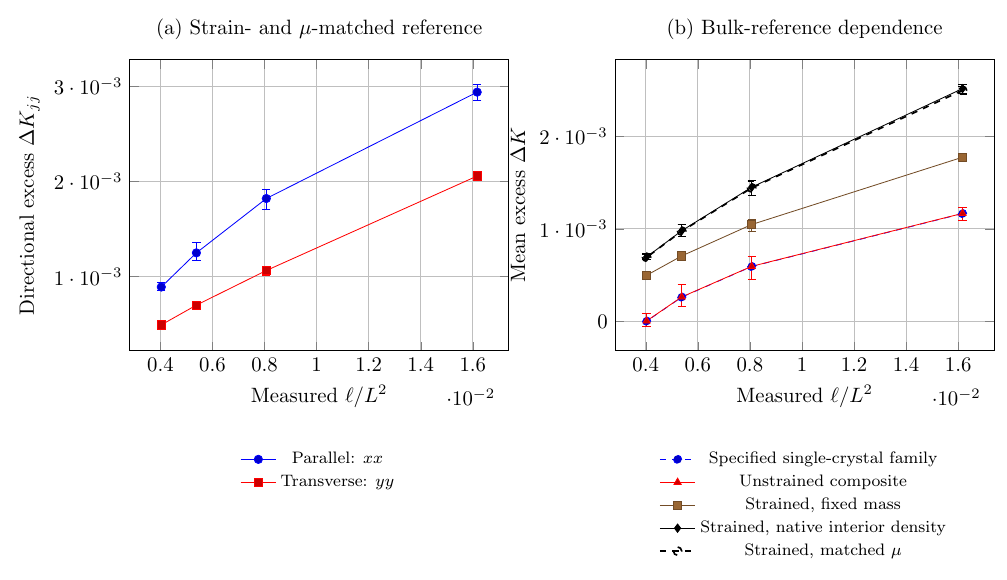}
\caption{Directional excess and bulk-reference dependence for thirty-two independently prepared bicrystals. Symbols show size means; error bars describe preparation resampling. Panel (a) uses the recommended strain- and chemical-potential-matched reference; both diagonal excesses are positive. Panel (b) includes that reference as a dashed black curve with open circles, nearly coincident with the native-density curve. Different bulk conventions define different excess observables.}\label{fig:interface-directional}
\end{figure*}

In the fixed-density ensemble, the excess relative to the prescribed crystal falls from $1.17\times10^{-3}$ at $L=128$ to $3\times10^{-6}$ at $L=512$, where it is compatible with zero. Relative to strained composites it remains positive at all sizes (Fig.~\ref{fig:interface-directional}).

Proportional fits $\Delta\overline K=\Gamma_K \nu$ give coefficients from 0.067 to 0.163 depending on the reference, and 0.162 for the recommended one (Table~\ref{tab:reference-scaling}). These are systematic changes of the comparison state, not random scatter around one material constant.
\begin{table*}[!t]\centering\small
\begin{tabular}{lrrr}\toprule
Reference & $\Gamma_K$ & RMS $(10^{-4})$ & Held-out RMS $(10^{-4})$\\\midrule
Prescribed crystal & 0.0674 & 2.11 & 2.48\\
Unstrained composite & 0.0676 & 2.10 & 2.47\\
Strained, fixed mass & 0.1155 & 1.14 & 1.94\\
Strained, native density & 0.1626 & 1.39 & 2.36\\
Strained, matched chemical potential & 0.1616 & 1.37 & 2.36\\
\bottomrule\end{tabular}
\caption{Proportional slopes for the same thirty-two states under different references. $\Gamma_K$ has units of density times length; at $n_0=1$ it equals the fraction-based slope numerically. RMS is the root-mean-square prediction residual in stiffness units; held-out RMS omits one size at a time.}\label{tab:reference-scaling}
\end{table*}

\section{Microscopic transport and reference sensitivity}\label{sec:mechanism}
The sensitivity to the bulk state is measurable without a defect. On the prescribed commensurate crystal branch, a one-percent isotropic dilation at fixed mean density reduces the half-trace stiffness by approximately 5--7\% over $80\leq\Lambda\leq128.68$; a one-percent compression increases it by approximately 6--9\%. Small differences in the comparator can therefore be comparable to the contribution assigned to a grain boundary (Appendix~\ref{app:transport-model}).

\begin{figure*}[!t]\centering\includegraphics[width=\textwidth]{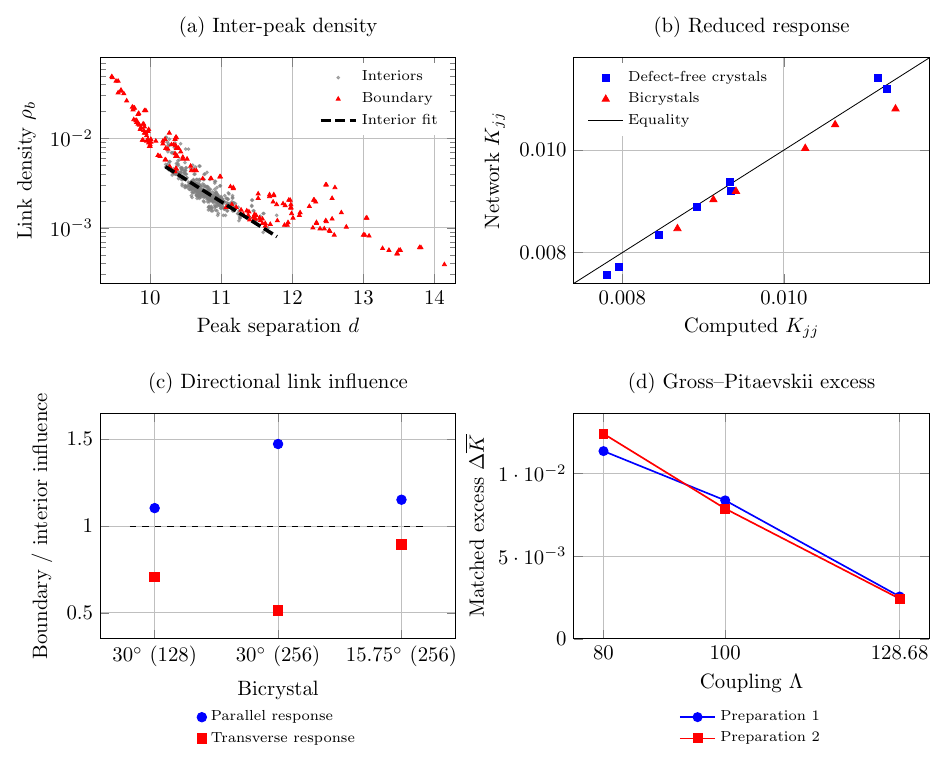}
\caption{Microscopic transport and reference-controlled response. (a) Inter-peak density in grain interiors and boundary regions; the line is the empirical interior fit over its sampled range. (b) Diagonal network and GP stiffnesses at $\Lambda=128.68$, with one calibrated coupling constant. (c) Ratio of the mean stiffness changes produced by halving equal numbers of boundary and interior couplings on each fixed network; labels give angle and box side. (d) GP excess relative to strain- and chemical-potential-matched composites for two paired preparations across interaction strengths. Connecting lines identify the same preparation branch.}\label{fig:link-network}
\end{figure*}

In the strongly modulated regime, inter-peak densities are more than three orders of magnitude below the peak density. We join neighboring peaks by periodic Delaunay bonds and define $\rho_b$ as the minimum density along each bond. Grain-interior and reference bonds admit the empirical representation
\begin{equation}
\ln(\rho_b/\rho_\ast)\simeq-d/\lambda,\qquad\lambda\simeq0.9\simeq0.11a,
\label{eq:link-law}
\end{equation}
where $d$ is the bond length (peak separation) and $\rho_\ast$ a fitted density prefactor, so the logarithm is dimensionless, over the sampled range $10.2\leq d\leq11.8$ (Fig.~\ref{fig:link-network}a). A squared-distance representation has similar predictive accuracy over this interval; Eq.~\eqref{eq:link-law} describes the observed sensitivity rather than an asymptotic decay law. Boundary bonds span a wider range of separations and include short, high-density connections.

Weak-link descriptions of modulated superfluids are established~\cite{Biagioni2024}. Here their spatial organization is examined with a network whose phase energy is
\begin{equation}
\mathcal E_{\rm net}(\boldsymbol q)=\frac{1}{2L^2}\min_{\{\varphi_s\}}\sum_bJ_b(\boldsymbol q\cdot\boldsymbol r_b+\varphi_t-\varphi_s)^2,\qquad J_b=c\rho_b.
\label{eq:network-energy}
\end{equation}
Here each bond $b=(s,t)$ is counted once, $\boldsymbol r_b$ points from peak $s$ to the appropriate periodic image of peak $t$, $\varphi_s$ are dimensionless periodic node phases with one fixed gauge, and $J_b$ has energy units. The calibration constant $c$ supplies the conversion from density to coupling. The network stiffness is $K^{\rm net}_{ij}=\partial_{q_i}\partial_{q_j}\mathcal E_{\rm net}|_{\boldsymbol q=0}$, in the same units as $K$. The constant $c$ is fixed by matching the half-trace stiffness of the prescribed $L=128$ defect-free crystal. The resulting network captures the dominant tensor response of the other crystals and bicrystals (Fig.~\ref{fig:link-network}b). It uses density profiles from the stationary GP solutions and is therefore a reduced description, not a prediction from peak positions alone. Small errors in total stiffness can become substantial after bulk subtraction, particularly for a weak transverse excess.

Changing selected couplings while keeping the geometry fixed tests which links control this reduced response. Capping compressed boundary links at the median interior coupling reduces both diagonal stiffnesses. To resolve their directional organization, equal numbers of randomly selected boundary or interior links are weakened by the same fraction. Boundary perturbations have a larger mean influence on parallel transport and a smaller influence on transverse transport in all three bicrystals (Fig.~\ref{fig:link-network}c). This supports a network interpretation of preferential flow along the boundary, without identifying a local coupling increase with an additive contribution to global stiffness.

The positive boundary excess also persists outside the original coupling. Two independently prepared bicrystals define paired branches at $\Lambda=80$, 100 and 128.68, with the same mean density and interaction range. All six states have positive diagonal excesses relative to their strain- and chemical-potential-matched composites (Fig.~\ref{fig:link-network}d). The smaller coupling reconstructs the crystal with fewer sites, so these results establish persistence of the effect rather than a comparison at identical interior conditions. The magnitude and anisotropy vary with coupling. The network approximation is also less accurate under transfer to weaker modulation, consistent with the importance of the full density profile beyond a single link minimum.

\section{Preparation variability and crystalline conditions}
\begin{figure*}[!t]\centering\includegraphics[width=\textwidth]{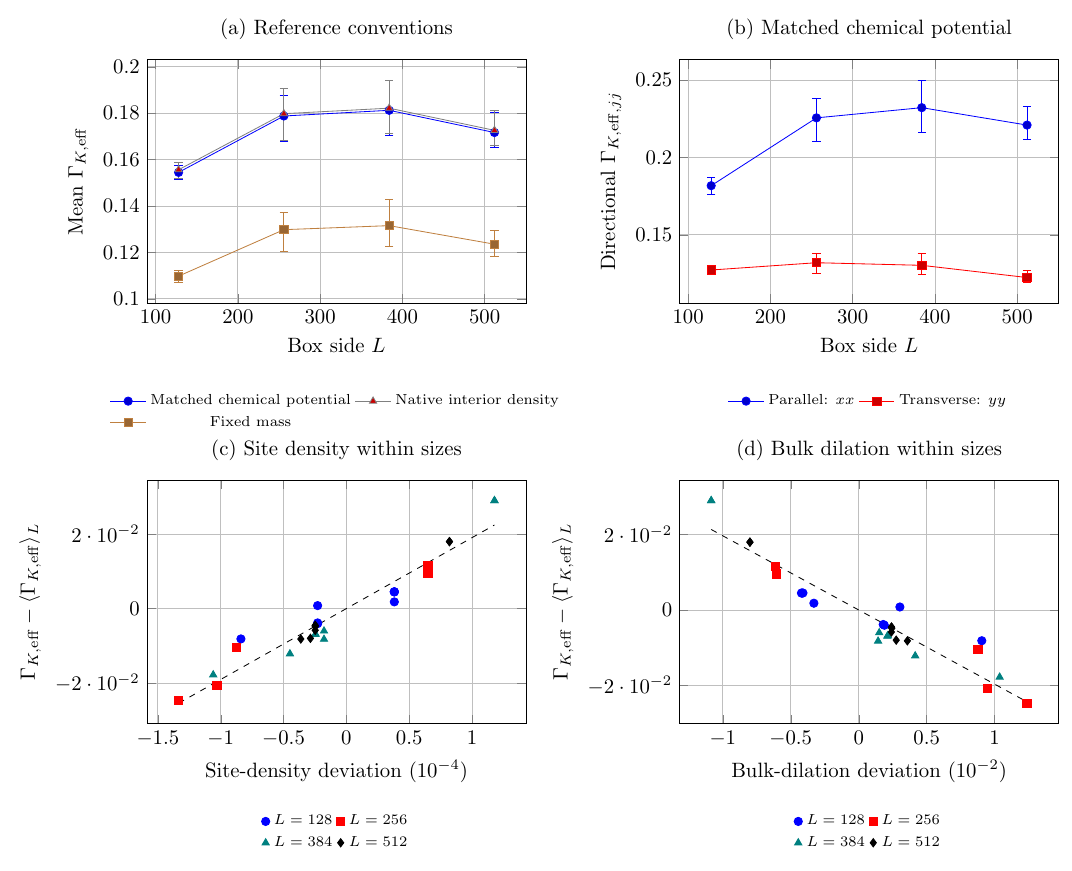}
\caption{Preparation-specific coefficients under different bulk conventions and their association with crystalline structure. Error bars show preparation resampling; correlations are evaluated within size.}\label{fig:effective-gamma}
\end{figure*}

For each preparation we define the effective coefficient
\begin{equation}
\Gamma_{K,\mathrm{eff},ij}=\frac{K^{\rm GB}_{ij}-K^{\rm bulk}_{ij}}{\ell/L^2}.
\end{equation}
Its half-trace is $\Gamma_{K,\mathrm{eff}}=(\Gamma_{K,\mathrm{eff},xx}+\Gamma_{K,\mathrm{eff},yy})/2$. With the recommended reference both diagonal excesses are positive, but the coefficients vary between preparations and sizes (Fig.~\ref{fig:effective-gamma}).

After subtracting size means, the half-trace coefficient correlates with the final site density ($r=0.965$), and a descriptor fit reduces the held-out-preparation RMS by 72\%. The correlation persists for native-density and fixed-mass references ($r=0.971$ and 0.969). It is consistent with Sec.~\ref{sec:mechanism}: at fixed area, more sites generally imply shorter bonds and stronger inter-peak connections. Site density and bulk dilation are nearly collinear ($r=-0.998$) and both are confounded with size, so their separate effects cannot be identified.

Numerical changes in the half-trace coefficient are at most $7\times10^{-5}$, compared with interface-definition and exclusion-depth sensitivities of $1.9\times10^{-3}$ and $3.0\times10^{-3}$. Preparation spread and reference choice dominate the variability.

\section{Interface scaling with comparable grain interiors}
\subsection{Interior-controlled ensemble}\label{sec:controlled}
To separate size from interior conditions, a commensurate family has $L=mL_0$, with integer $m$, $L_0=135.019$ and $N_g=288m$, giving $h=0.4688$. Its lattice spacing minimizes the energy along the specified lattice family. With $U_0$ and $a$ fixed, the mean density $n_0$ is adjusted so that the grain-interior density is one. The physical field $\psi$ and the unit-mean-density field $\phi$ satisfy $\psi=\sqrt{n_0}\,\phi$; $\phi$ is computed at coupling $\Lambda_\phi=n_0\Lambda_0$, $f_\phi$ denotes its tensor from Eq.~\eqref{eq:tensor}, and the physical stiffness is $K=n_0 f_\phi$.

Of twenty-eight independent stationary states, twelve are mutually comparable across three sizes: four at $L=270.04$, three at $405.06$ and five at $540.08$. Matching uses interior density, strain, cell area, orientation and homogeneous-cell response, together with chemical potential and grain fraction, but not the defective response. Within this group, the largest differences are $0.0038$ in a Green-strain diagonal, $2\times10^{-4}$ in grain density, $9\times10^{-4}$ in chemical potential and $4\times10^{-5}$ in a bulk-tensor component (Appendix~\ref{app:matching}).

\subsection{Finite-range coefficient and size residuals}
With the recommended composite, all twelve half-trace excesses are positive. The proportional slope is $\Gamma_K=0.1127$, in units of density times length, with RMS residual $7.3\times10^{-6}$. Parallel and transverse slopes are 0.132 and 0.093 (Table~\ref{tab:matched-law}), consistent with the parallel channels of Sec.~\ref{sec:mechanism}.

Residual bulk mismatch does not change this result. Replacing each state's grain tensors by those of every other comparable state, at fixed defective response, interface length and partition, gives 288 comparisons, all with positive half-trace and diagonal excesses. The largest half-trace shift is 3.8\% at the donor chemical potential and 1.4\% at the recipient one; tensor-order bounds over the whole library also remain positive (Appendix~\ref{app:bulk-robustness}).

\begin{table*}[!t]\centering\small\begin{tabular}{lrr}\toprule Component & Fitted $\Gamma_K$ & Range of size-mean $\Gamma_{K,\rm eff}$\\\midrule
Half trace & 0.1127 & [0.1107, 0.1135]\\
Parallel & 0.1322 & [0.1311, 0.1326]\\
Transverse & 0.0932 & [0.0902, 0.0944]\\
\bottomrule\end{tabular}

\caption{Physical interface coefficients for the twelve mutually comparable preparations. Ranges are differences between size means, not confidence intervals. Units are density times length.}\label{tab:matched-law}
\end{table*}

\begin{figure*}[!t]\centering\includegraphics[width=\textwidth]{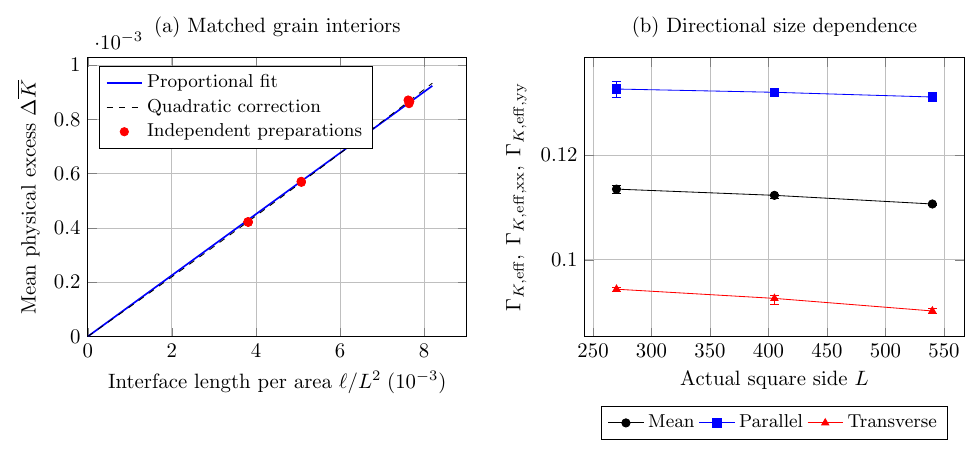}
\caption{Interface scaling with comparable grain interiors. (a) Individual physical excesses with proportional and quadratic fits. (b) Directional size-mean effective coefficients; error bars are resampling summaries over three to five states per size.}\label{fig:matched-law}
\end{figure*}

Size-mean coefficients decrease from 0.1135 to 0.1123 and 0.1107, a 2.5\% range, and two of the three conditional held-out-size residual intervals exclude zero after Bonferroni adjustment. The slope therefore summarizes the measured range without establishing strict size independence; a power fit gives an exponent of 1.03, over only a factor of two in interface density. In a larger box, $L=675.09$, three stationary states with interiors comparable to the $L=540.08$ group give prediction errors of $-2.5\%$ for the mean and $-4.5\%$ for the transverse component, without refitting (Appendix~\ref{app:validation}).

\subsection{Independent interface-length changes at fixed area}
At fixed $L=405.06$, eight independent blocks each pair freely relaxed states with two and four interfaces; all sixteen are stationary and keep their interface counts. In the four pairs whose interiors satisfy the matching criteria, adding interface length increases the bulk-subtracted response. With $\delta\nu=\nu_4-\nu_2$ and $\delta K=\Delta\overline K_4-\Delta\overline K_2$, the paired slope is
\begin{equation}
\Gamma_K^{\rm pair}=\frac{\sum_p\delta\nu_p\,\delta K_p}{\sum_p\delta\nu_p^2}=0.1162,
\label{eq:paired-slope}
\end{equation}
where $p$ labels accepted pairs and the subscripts 2 and 4 denote interface counts, with individual slopes between 0.1136 and 0.1183 (coefficient of variation 1.8\%). This is 3.1\% above the three-size coefficient, independent support for the finite-range response to added length. Only two of the blocks are also comparable with each other, so constancy of the coefficient across crystal conditions is not tested (Fig.~\ref{fig:paired-length}, Appendix~\ref{app:paired-length}).

\section{Discussion}
The grain-boundary contribution to superfluid stiffness is an excess whose value depends on the bulk state subtracted. The Gibbs--Cahn construction~\cite{Gibbs1878,Cahn1979} clarifies the roles of the bicrystal and its comparator without supplying a thermodynamic identity: phase relaxation throughout the sample prevents an assumed decomposition into a unique additive line contribution. Controlled deformations of defect-free crystals show that modest differences in spacing produce stiffness changes comparable to the boundary excess. Matching the comparator to the measured interior strain and chemical potential controls this source of ambiguity.

The qualitative enhancement and the length-per-area description of Ref.~\cite{Sepulveda2010} are supported under specified crystal conditions. The link network connects the directional response to the organization of boundary connections, but its errors after bulk subtraction prevent a quantitative identification of the excess with compressed bonds alone. Positive reference-controlled excesses at three couplings extend the result beyond the original interaction strength. Their differing crystal structures and the distinct slopes of the fixed-density and interior-controlled families show why a common material coefficient cannot be inferred by pooling these states.

An independent implementation~\cite{Blakie2024} agrees with our primitive-cell responses at its quoted precision (Appendix~\ref{app:external-benchmark}); its Fig.~5(a) also shows lower responses than those of Ref.~\cite{Sepulveda2010}. The perfect-crystal value 0.051 of Ref.~\cite{Sepulveda2010} is not recovered by our converged $L=256$ reference, 0.00913. Incompletely relaxed states can give responses of that order (0.0545 before convergence to 0.0078 in a primitive cell), which offers a possible explanation (Appendix~\ref{app:baseline}). The absolute difference does not affect the comparisons within this study.

The interface-scaling ensembles use $\Lambda_0=\pi a^4U_0=128.68$, while the paired coupling comparisons establish persistence of the sign of the excess over a broader interval. They do not determine a coupling-independent line coefficient. Soft-core interactions are motivated by Rydberg dressing~\cite{Henkel2010,Cinti2010,Macri2013}, and finite-temperature phase diagrams~\cite{Peotta2025} complement these zero-temperature results. The coefficients are not calibrated to dipolar experiments~\cite{Biagioni2024}, and the preparation ensembles describe variation under specified initial conditions rather than thermal probabilities.

\section{Conclusions}
The contribution assigned to a grain boundary requires an explicitly specified bulk response. Reference sensitivity can be comparable to the inferred enhancement, and controlled lattice deformations demonstrate that this ambiguity has a substantial physical scale. A strain- and chemical-potential-matched composite supplies an operational comparator. Relative to it, the boundary excess remains positive across comparable preparations and across the three interaction strengths studied.

Within the interior-matched family, the excess follows an approximate length-per-area relation with coefficient 0.113, supported by an independent fixed-area value of 0.116. These are conditional coefficients: the directional response requires a tensorial description, and finite-size residuals and differences between crystal families limit a common-coefficient interpretation. Reduced link networks explain aspects of the spatial organization of transport but do not replace the full response calculation for small excesses. The resulting description quantifies the known grain-boundary enhancement while separating it from changes in the surrounding crystal.

\begin{acknowledgments}
%The author thanks Sergio Rica and Christophe Josserand for introducing the problem studied here. This research received no specific funding.
\end{acknowledgments}

\section*{Conflict of interest}
The author declares no competing interests.

\section*{Data availability}
The stationary fields, parameter files, response tensors, and analysis code supporting the findings of this article are available from the author upon reasonable request.

\clearpage
\appendix
\section{Preparation and structural measurements}\label{app:preparations}
\subsection{Periodic seeds and reference branches}\label{sec:preparations}
A commensurate seed is constructed from integer reciprocal vectors $\boldsymbol q_1$ and $\boldsymbol q_2$, in units $2\pi/L$:
\begin{equation}
 \begin{aligned}
 u_{\mathrm{seed}}\propto{}&1+\epsilon\big[\cos(\boldsymbol Q_1\cdot\boldsymbol x)
 +\cos(\boldsymbol Q_2\cdot\boldsymbol x)\\
 &+\cos((\boldsymbol Q_2-\boldsymbol Q_1)\cdot\boldsymbol x)\big],\\
 \boldsymbol Q_i={}&\frac{2\pi}{L}\boldsymbol q_i,
 \end{aligned}
 \label{eq:seed}
\end{equation}
with $\epsilon=0.1$ followed by normalization. The corresponding direct-cell matrix is $B=L(\boldsymbol q_1,\boldsymbol q_2)^{-\mathsf T}$. The determinant $|\det(\boldsymbol q_1,\boldsymbol q_2)|$ gives the seeded cell count; the final number of maxima $N_s$ is measured independently. The square torus requires strain for an exactly periodic triangular pattern. A seed designation is therefore neither a certificate of hexagonal coordination nor evidence of a globally minimizing crystal.

The $30^\circ$ family uses grain-A vectors $(0,14)$ and $(12,7)$ at $L=128$. Grain B interchanges their Cartesian components, giving $(14,0)$ and $(7,12)$. Both sets are doubled at $L=256$, preserving seeded spacing and strain. Two orientations are joined across $y=L/2$. Periodicity creates two boundaries, including the seam of the box. The nominal angle describes the seed; the final structure is measured.

For the second family at $L=256$, the seed vectors are
\begin{align}
 A:&\quad \boldsymbol q_1=(0,28),\quad \boldsymbol q_2=(24,14),\\
 B:&\quad \boldsymbol q_1=(-28,-7),\quad \boldsymbol q_2=(-20,19).
\end{align}
Both determinants have magnitude 672. Their nominal reciprocal misorientation is $15.535^\circ$. The final bicrystal has 644 peaks. We therefore also prepare 644-site single crystals with bases $(0,28),(23,14)$ and $(-28,-7),(-20,18)$.

\subsection{Independent preparations and interface length}
At fixed mean density one, we prepared eight independent bicrystals in each of four square boxes, $L=128,256,384,512$, with $a=8$ and $\Lambda=128.68$. Equation~\eqref{eq:seed} has amplitude $\epsilon=0.1$ and the reciprocal vectors of Sec.~\ref{sec:preparations} multiplied by $m_L=L/128$: grain A uses $(0,14m_L),(12m_L,7m_L)$ and grain B interchanges Cartesian components. Random grain translations, boundary positions and roughness phases define the preparation ensemble. The two boundaries initially share a displacement $(a/4)\sin(2\pi x/L+\gamma_1)+(a/8)\sin(4\pi x/L+\gamma_2)$. The added perturbation is $0.001\sum_{s=1}^{8}\cos(\boldsymbol Q_s\cdot\boldsymbol x+\chi_s)/2$, where $\boldsymbol Q_s=(2\pi/L)(q_{xs},q_{ys})$, integer components satisfy $|q_{is}|\leq\max(2,L/32)$ and the zero vector is excluded. The phases $\gamma_1$, $\gamma_2$ and $\chi_s$ are uniform on $[0,2\pi)$. The random seed is $20261018+100003L+1009\kappa$, with preparation index $\kappa=0,\ldots,7$. Final site counts and strain are measured independently.

We trace periodic zero contours of a Gaussian interpolation of $\operatorname{Re}\psi_6$, with width $a$ and sampling $a/4$. Two winding boundaries must carry more than 95\% of contour length, both grains must have at least twelve ordered interior sites, and the misorientation must lie between $27^\circ$ and $30^\circ$. Extraction is repeated at sampling $a/2$ and widths $0.75a$ and $1.25a$. All thirty-two preparations meet this structural class. Each is measured at $h=0.5$ and $R\leq10^{-9}$; preparations zero and one at each size receive $h=0.25$ continuations.

\subsection{Coordination and local enrichment}

Density maxima are detected periodically, refined locally, and used to construct periodic Voronoi coordination numbers $z_s$. We record the site count $N_s$, the number $N_{\ne6}$ of sites with $z_s\ne6$, and the torus coordination charge $\sum_s(6-z_s)$. The latter vanishes for all retained configurations. It is a consistency check, not a vacancy count or a Burgers-vector measurement.

The local orientational order is
\begin{equation}
 \psi_6(s)=\frac{1}{z_s}\sum_{t\in\mathcal N_s}\exp(6i\vartheta_{st}).
\end{equation}
Here $i$ in the exponential is the imaginary unit, $\mathcal N_s$ is the set of Voronoi neighbors of site $s$ and $\vartheta_{st}$ the angle of the bond from $s$ to $t$.
Bulk orientations are obtained from the argument of the mean $\psi_6$ among sixfold sites in the strips $y/L\in(0.125,0.375)$ and $(0.625,0.875)$. The orientation is one sixth of that argument, modulo $60^\circ$; misorientation is the smallest angular difference modulo $60^\circ$. Both the magnitude of that average and the site count are retained. This estimator is descriptive when a reconstructed interface curves or approaches a sampling strip.

The continuous cell mass $M_s=\int_{\mathcal V_s}\rho\,\dd^2x$, with $\mathcal V_s$ the Voronoi cell of site $s$, is assigned by nearest-peak periodic Voronoi partitioning of grid points. Its coefficient of variation measures mass redistribution among density peaks. It does not measure integer atomic occupation or its quantum fluctuations.

\begin{table*}[!t]\centering\small
\begin{tabular}{rrrrrr}
\toprule
Nominal angle ($^\circ$) & $L$ & $N_{\ne6}$ & $100p_{\mathcal M}$ & $\mathcal Q_x$ & $\mathcal Q_y$\\
\midrule
30 & 128 & 28 & 26.35 & 1.301 & 0.730\\
30 & 256 & 56 & 13.08 & 1.582 & 0.530\\
15.535 & 256 & 32 & 8.55 & 1.266 & 0.844\\
\bottomrule\end{tabular}

\caption{Directional stiffness enrichment near coordination defects. Angles label nominal seeds; all states use $h=0.5$. The nominal interfaces are parallel to $x$.}\label{tab:local}
\end{table*}
\section{Numerical convergence and response validation}\label{app:numerics}

Fourier transforms evaluate the kinetic operator and disk convolution using Eq.~\eqref{eq:kernel}. Fields are minimized at fixed norm with projected, preconditioned nonlinear conjugate gradients, line search, and a Newton--CG refinement near stationarity. The residual is defined in Eq.~\eqref{eq:residual}.
The minimizer step is an optimization parameter, not a physical relaxation time. Norm conservation or a small energy decrement alone is not used as a convergence criterion. A small residual does not establish a local or global energy minimum.

Limited-memory quasi-Newton (L-BFGS) iterations~\cite{LiuNocedal1989} complement the minimization where conjugate-gradient progress is slow. The response equations are solved by projected conjugate gradients with a multigrid preconditioner~\cite{Briggs2000}, with target relative residual $10^{-9}$ and explicit recomputation of the final residual. Tensor reciprocity and eigenvalues in $[0,1]$, up to numerical tolerances, are checked. Collocation derivatives annihilate the real Nyquist mode consistently in the boost, momentum, and response. No explicit dealiasing filter is used; refinement and the spectral tail provide numerical diagnostics.

All states entering the ensemble response analysis satisfy $R\leq10^{-9}$. The individual branch examples use $R\leq10^{-8}$, except the coarse low-angle preparation, accepted at $R\leq10^{-7}$ with achieved residual about $4.4\times10^{-8}$. Its refined state is converged to $10^{-10}$. Independent finite-boost checks use residuals below $10^{-12}$. The ensemble statistics use the stricter tolerance.

\subsection{Numerical verification}\label{sec:numerical-verification}
Stationarity residuals, grid refinement, tensor identities, independent primitive-cell benchmarks, and reproduction from archived fields test the numerical calculations. Figures render computed fields and tabulated data.

\subsection{External triangular-cell benchmark}\label{app:external-benchmark}
We compare with the triangular stationary states in the discussion accompanying Figs.~4 and 5(a) of Blakie~\cite{Blakie2024}. Its coupling $\Lambda=\pi nma^4U_0/\hbar^2$ agrees with Eq.~\eqref{eq:kernel}. At each coupling we minimize the energy over the equilateral primitive-cell side $b$, with vectors $(b,0)$ and $(b/2,\sqrt3b/2)$, at fixed mean density one. An initial scan over $b/a\in[1.25,1.55]$ brackets an interior energy minimum, followed by a golden-section search on the $32^2$ grid. The final side bracket has width below $5\times10^{-7}$ in $b/a$. A triangular seed of amplitude 0.4 retains the crystalline branch near the first-order transition; each optimized crystal has lower energy than the uniform state. The response is then recomputed independently on $32^2$, $64^2$ and $128^2$ grids, with stationary residuals below $10^{-11}$.
\begin{table*}[!t]\centering\small
\begin{tabular}{rrrr}\toprule
$\Lambda$ & Optimized $b/a$ & Present $\fbar$ & Blakie, quoted $f_s$\\\midrule
40 & 1.513 & 0.6321 & 0.63\\
100 & 1.422 & 0.02346 & 0.02\\
\bottomrule\end{tabular}
\caption{External primitive-cell benchmark against the values explicitly quoted in Ref.~\cite{Blakie2024}, in the discussion of Fig.~4. The last column retains the precision of that source; it is not a set of high-precision reference data.}\label{tab:external-benchmark}
\end{table*}
Both results agree with the quoted values when rounded to two decimal places. Across the three grids, changes in the half-trace response and diagonal anisotropy are below $10^{-10}$. This checks the absolute response against an independent implementation in weakly and strongly modulated triangular states. The rounded external values do not establish a high-precision relative error, especially for the smaller response at $\Lambda=100$; mesh agreement is an internal numerical check rather than an uncertainty on the external comparison.

\subsection{Spatial and boost convergence}
Table~\ref{tab:spatial} holds the $L=128$, 168-site single-crystal branch fixed while refining the grid. The $h=2$ result is inadequate. The two eigenvalues at $h=0.5$ agree with their $h=0.25$ counterparts to the recorded precision. The relative difference of order $10^{-11}$ is a comparison between these runs, a check on this branch, with separate defective-state checks below.
\begin{table*}[!t]\centering\small
\begin{tabular}{rrrrr}
\toprule
$N_g$ & $h$ & $f_1$ & $f_2$ & Relative difference\\
\midrule
64 & 2.00 & 0.013167908 & 0.028464558 & $2.05\times10^{0}$\\
128 & 1.00 & 0.008927631 & 0.009330980 & $3.61\times10^{-5}$\\
256 & 0.50 & 0.008927953 & 0.009331078 & $1.92\times10^{-11}$\\
512 & 0.25 & 0.008927953 & 0.009331078 & 0\\
\bottomrule\end{tabular}

\caption{Grid refinement for one 168-site branch at $L=128$. The last column is the maximum relative eigenvalue difference from $N_g=512$. Zero denotes the reference entry.}\label{tab:spatial}
\end{table*}

For a defect-free branch, the linear response is independently compared with
\begin{equation}
 f^{\mathrm s}_{xx}(v)=1-\frac{P_x(+v)-P_x(-v)}{2vn_0}.
\end{equation}
Here $P_x(\pm v)$ is evaluated with boost $\boldsymbol v=\pm v\hat{\boldsymbol x}_x$; this central-difference estimate tends to $f^{\mathrm s}_{xx}$ as $v\to0$. The absolute discrepancies for $v=10^{-3}$, $5\times10^{-4}$, and $2.5\times10^{-4}$ are respectively $1.22\times10^{-7}$, $2.54\times10^{-8}$, and $4.04\times10^{-9}$. Smaller boosts require correspondingly tighter stationary solutions; insufficient tolerance can increase the error through division by $v$. We do not assign an exact asymptotic order to the last points.

For the low-angle bicrystal, tightening $R$ from $10^{-8}$ to $10^{-10}$ changes the largest tensor component difference by $\LowAngleToleranceChange$. The final residual is $\LowAngleResidual$. Between the coarse preparation and the tightly converged fine state the largest component change is $\LowAngleGridChange$, while the peak and non-sixfold counts remain 644 and 32. These checks support the quoted comparisons at the studied resolution, but do not establish an error bound for every local field.

\begin{figure*}[!t]\centering\includegraphics[width=\textwidth]{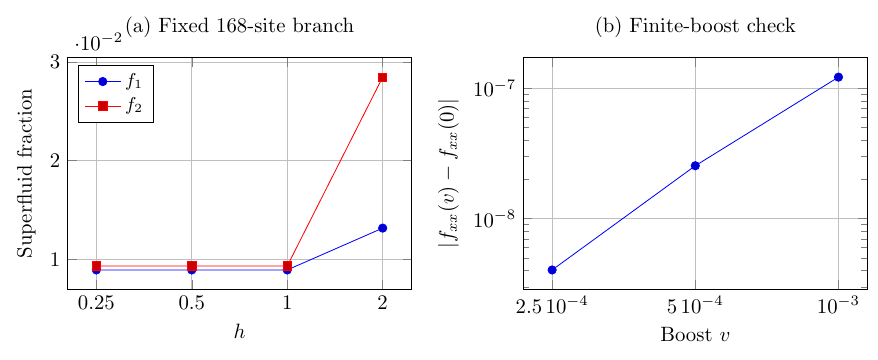}
\caption{Numerical validation: (a) eigenvalues under grid refinement on the fixed 168-site branch; (b) central finite-boost discrepancies relative to linear response.}\label{fig:validation}
\end{figure*}

Three paired continuations of the interior-controlled ensemble halve $h$ from 0.468815 to 0.234408 at fixed physical parameters. Maximum tensor changes range from $2.12\times10^{-10}$ to $4.53\times10^{-10}$. Recomputing the bulk subtraction changes the half-trace excess by at most $2.60\times10^{-8}$.

\section{Branch and defect controls}\label{app:controls}
\subsection{A size-dependent excess in the \texorpdfstring{$30^\circ$}{30-degree} family}
Table~\ref{tab:size} gives the two-size comparison at $h=0.5$. The response increases by $13.85\%$ at $L=128$ and $8.23\%$ at $L=256$. The smaller-box enhancement persists under further refinement to $h=0.25$. However, the bicrystals contain 161 and 658 sites, compared with 168 and 672 in the references. The density of crystalline sites is not fixed by fixing total mass.
\begin{table*}[!t]\centering\small
\begin{tabular}{rrrrrrr}
\toprule
$L$ & $N_g$ & $f_{\rm ref}$ & $\fbar_{\rm GB}$ & $\mathcal H$ (\%) & $N_s^{\rm ref}/N_s^{\rm GB}$ & $N_{\ne6}$\\
\midrule
128 & 256 & 0.00912952 & 0.01039405 & 13.851 & 168/161 & 28\\
256 & 512 & 0.00912955 & 0.00988073 & 8.228 & 672/658 & 56\\
\bottomrule\end{tabular}

\caption{Specified $30^\circ$ family. $N_s^{\rm ref}/N_s^{\rm GB}$ lists the measured site counts. All rows use $h=0.5$.}\label{tab:size}
\end{table*}

The absolute excess decreases from $1.26453\times10^{-3}$ to $7.51184\times10^{-4}$. The products $L\Delta f$ are $0.16186$ and $0.19230$, differing by approximately $19\%$ relative to the smaller-box value. The decrease is compatible with a boundary contribution becoming less important in larger domains, but two sizes do not validate $\Delta f\propto L^{-1}$ or determine an exponent.

\begin{figure*}[!t]\centering
\includegraphics[width=.96\textwidth]{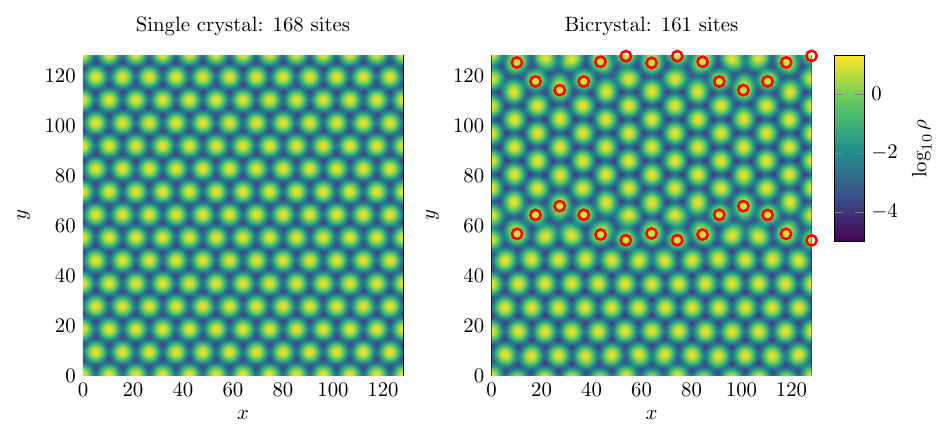}
\caption{Density patterns at $L=128$, $N_g=256$, with a common logarithmic scale. The left panel is the 168-site reference; the right is the 161-site bicrystal. Red circles mark sites with $z_s\ne6$. The two interface regions are displaced from their initial positions, which motivates masks based on the final structure.}\label{fig:density}
\end{figure*}

\subsection{Sparse coordination defects and density depletion}
A separate random preparation at $L=128$ ends with 146 peaks and four anomalous sites, two with coordination five and two with coordination seven. A defect-free 146-site control with reciprocal basis $(5,12),(13,2)$ has $\fbar=\SparseControlMean$, compared with $\SparseDefectMean$ for the defective realization, a relative difference of $\SparseEnhancement\%$. This is one realization. The coordination pairs do not by themselves determine Burgers vectors; strain and orientation also differ between the compared states.

To distinguish a continuous density perturbation from a quantum occupation defect, we also modify a converged 168-site crystal through
\begin{equation}
 \begin{gathered}
 u'(\boldsymbol x)=C u(\boldsymbol x)\sqrt{1-D\exp[-r_{\rm per}^2/(2w^2)]},\\
 w=a/4,\quad D\in\{0.5,1\},
 \end{gathered}
\end{equation}
where $r_{\rm per}$ is the periodic distance from a selected central maximum and $C$ restores the total norm. These operations remove continuous density locally and redistribute it through normalization; they do not remove a specified integer number of atoms.

The initial cell-mass coefficients of variation are $0.02937$ and $0.05889$. Both preparations relax to 168 peaks with sixfold coordination. Their final coefficients are approximately $1.61\times10^{-5}$, close to the control value $1.69\times10^{-5}$, and their mean responses differ from the control by at most $1.86\times10^{-9}$. The preparations thus leave no detectable stationary structural vacancy under this protocol. No physical healing time or defect lifetime follows from the optimization iterations. The result is not a test of the occupation-defect mechanism of Ref.~\cite{Cinti2014}, which requires many-body occupation fluctuations absent from a single coherent GP field.

\section{Preparation statistics and model sensitivity}\label{app:statistics}

\subsection{Robustness to residual bulk mismatch}\label{app:bulk-robustness}
The twelve mutually comparable states supply a finite library of grain-interior deformations. For each recipient, its physical defective tensor, measured interface length and grain partition are fixed. Each donor supplies both homogeneous grain tensors, either at its own chemical potential or after density readjustment to the recipient chemical potential. This gives $12\times12\times2=288$ comparisons, without introducing additional independent preparations. The self-reference exactly reproduces the original bulk subtraction at recorded precision.

\begin{table*}[!t]\centering\small
\begin{tabular}{lrrr}\toprule
Chemical-potential convention & Maximum excess shift & Minimum $\Delta\overline K$ & Slope envelope\\\midrule
Donor & 3.84\% & $4.0543\times10^{-4}$ & [0.10795, 0.11745]\\
Recipient & 1.41\% & $4.1983\times10^{-4}$ & [0.10829, 0.11720]\\
\bottomrule\end{tabular}
\caption{Sensitivity to the observed interior library. Excess shifts are relative to each recipient's original excess. Slope envelopes cover independent grain-library substitutions and convex tensor mixtures on fixed partitions; they are not confidence intervals.}\label{tab:bulk-robustness}
\end{table*}

For each recipient, let $K_g^0$ be its original homogeneous tensor in grain $g$. The extreme generalized eigenvalues of all donor tensors define common bounds $0<\alpha\leq\beta$ over both grain libraries such that $\alpha K_g^0\preceq K_g\preceq\beta K_g^0$. The variational phase energy preserves this matrix order under composition:
\begin{equation}
\alpha K_{\rm bulk}^0\preceq K_{\rm bulk}\preceq\beta K_{\rm bulk}^0,
\end{equation}
where $A\preceq B$ denotes the Loewner order, i.e., $B-A$ is positive semidefinite.
Consequently $K_{jj}^{\rm GB}-\beta K_{{\rm bulk},jj}^0$ bounds each excess from below for arbitrary independent choices from the two grain libraries and their convex tensor mixtures. The smallest half-trace lower bounds are $3.7685\times10^{-4}$ and $3.9885\times10^{-4}$ for donor and recipient chemical potentials, respectively; all diagonal lower bounds are positive. Because $\nu>0$, propagating these bounds through the proportional fit gives the slope envelopes in Table~\ref{tab:bulk-robustness}.

These are deterministic sensitivities to observed, jointly comparable interiors. They do not cover unseen tensor fields, changes in grain geometry or the full multidimensional matching-tolerance region, and they do not establish a universal coefficient.

\subsection{Preparation resampling and model comparisons}
SD denotes their sample standard deviation, and $r$ denotes the Pearson correlation. For $\nu=\ell/L^2$, the fitted half-trace models are $\Gamma_K\nu$, $b_0+\Gamma_K\nu$ and $\Gamma_K\nu+C_2\nu^2$, with intercept $b_0$ in density units and quadratic coefficient $C_2$ in density-times-length-squared units. Power fits use $A_p\nu^p$, with exponent $p$ and amplitude $A_p$ in density-times-length-to-the-power-$p$ units. Fits weight preparations equally. Four thousand bootstrap samples resample preparations within sizes. Omitting an entire size tests extrapolation across the measured range; paired numerical continuations are kept separate. Conditional percentile intervals are descriptive resampling summaries, not total uncertainty on a material constant.
\begin{table*}[!t]\centering\scriptsize
\begin{tabular}{rrrrrrr}\toprule
$L$ & $N_{\rm prep}$ & $\langle\ell\rangle$ & $\langle\Delta\overline K\rangle$ & Preparation SD & Resampling interval & Numerical change\\
 & & & $(10^{-3})$ & $(10^{-4})$ & $(10^{-3})$ & $(10^{-10})$\\\midrule
128 & 8 & 264.9 & 1.165 & 1.08 & [1.096, 1.233] & 3.08\\
256 & 8 & 528.8 & 0.5932 & 2.09 & [0.4507, 0.7029] & 3.02\\
384 & 8 & 793.6 & 0.2623 & 1.81 & [0.1564, 0.3977] & 5.22\\
512 & 8 & 1058 & 0.003234 & 0.994 & [$-0.05250$, 0.07954] & 5.23\\
\bottomrule\end{tabular}

\caption{Independent fixed-density ensemble. $N_{\rm prep}$ counts independent preparations; standard deviations describe their spread. Intervals are percentile resampling diagnostics conditional on the prescribed-crystal reference; paired numerical changes are reported separately.}\label{tab:ensemble}
\end{table*}

Under the prescribed-crystal reference the proportional slope is $0.06737$, with in-sample RMS $2.107\times10^{-4}$ and held-out-size RMS $2.484\times10^{-4}$. The corresponding held-out RMS values with an intercept and quadratic correction are $3.567\times10^{-4}$ and $1.012\times10^{-3}$. None establishes an asymptotic law.

\begin{figure*}[!t]\centering\includegraphics[width=\textwidth]{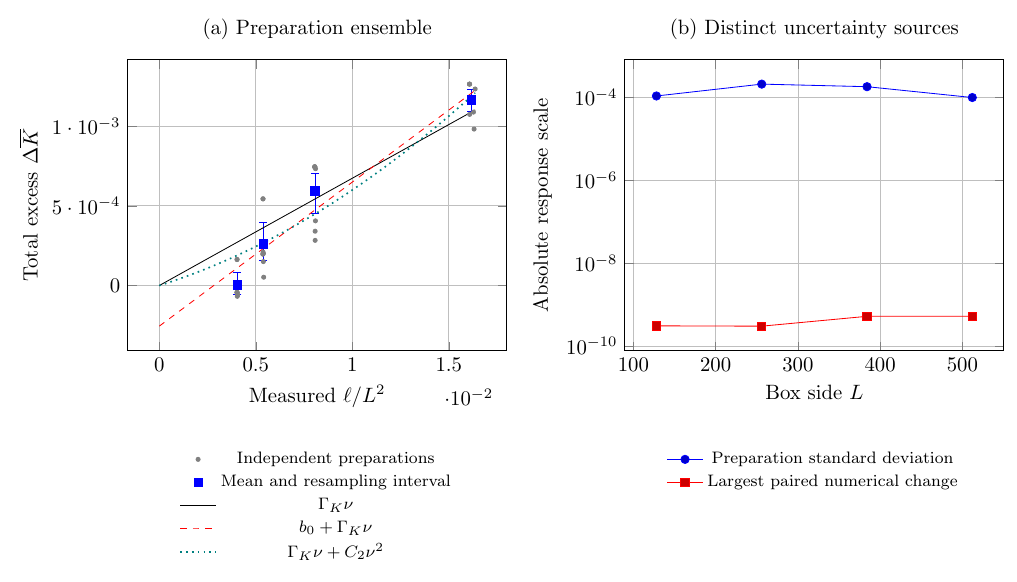}
\caption{Interface-length fits and preparation variability in the fixed-density ensemble. Points are independent preparations; squares show size means and conditional resampling intervals. Curves compare proportional, intercept and quadratic models.}\label{fig:interface-scaling}
\end{figure*}

\begin{table*}[!t]\centering\scriptsize
\begin{tabular}{rrrrr}\toprule $L$ & $\langle\Gamma_{K,\mathrm{eff}}\rangle$ [Resampling interval] & $\langle\Gamma_{K,\mathrm{eff},xx}\rangle$ & $\langle\Gamma_{K,\mathrm{eff},yy}\rangle$ & SD$(\Gamma_{K,\mathrm{eff}})$\\\midrule
128 & 0.15449 [0.15124,0.15745] & 0.18177 & 0.12721 & 0.00481\\
256 & 0.17879 [0.16771,0.18763] & 0.22564 & 0.13193 & 0.01593\\
384 & 0.18124 [0.17058,0.19408] & 0.23224 & 0.13025 & 0.01828\\
512 & 0.17166 [0.16526,0.18027] & 0.22098 & 0.12235 & 0.01118\\
\bottomrule\end{tabular}

\caption{Preparation-specific effective coefficients for the unit-density ensemble, with stiffness coefficients in density-times-length units (numerically equal to fraction-based coefficients at $n_0=1$). SD is preparation spread; intervals are conditional percentile resampling diagnostics.}\label{tab:effective-gamma}
\end{table*}

The nine descriptors use grain-area weights $\omega_A$ and $\omega_B=1-\omega_A$ and area-weighted interior strain means $E_{ij,g}$ and standard deviations $\sigma_{ij,g}$. They are site density $N_s/L^2$, interior density $\sum_g\omega_g n_g$, dilation $\sum_g\omega_g\operatorname{tr}E_g$, strain anisotropy $\sum_g\omega_g\sqrt{(E_{xx,g}-E_{yy,g})^2+4E_{xy,g}^2}$, strain heterogeneity $[\sum_g\omega_g(\sigma_{xx,g}^2+\sigma_{yy,g}^2+2\sigma_{xy,g}^2)]^{1/2}$, relative excess length $\ell/(2L)-1$, coordination anomalies per length $N_{\ne6}/\ell$, grain-area imbalance $|\omega_A-1/2|$, and misorientation. These densities are measured native-interior densities, independent of the density convention used for bulk subtraction. We analyze each descriptor separately after subtracting its size mean. Four thousand paired bootstrap samples and ten thousand joint tensor-outcome permutations preserve the size strata. Holm adjustment covers all twenty-seven descriptor/component tests. The half-trace coefficient correlates strongly with site density ($r=0.965$); held-out-preparation RMS decreases by 72.1\% relative to the training mean at the same size. Site density and bulk dilation are almost collinear ($r=-0.9983$), preventing independent causal attribution.

\begin{table*}[!t]\centering\scriptsize
\begin{tabular}{lrrrr}\toprule Descriptor & Within-size $r$ & Resampling interval & Holm $p_{\rm MC}$ & RMS reduction\\\midrule
Final site density & 0.965 & [0.943,0.983] & $0.0027^{\dagger}$ & 72.1\%\\
Interior density & 0.901 & [0.827,0.959] & $0.0027^{\dagger}$ & 53.4\%\\
Mean bulk dilation & -0.953 & [-0.981,-0.926] & $0.0027^{\dagger}$ & 67.9\%\\
Bulk strain anisotropy & -0.959 & [-0.982,-0.924] & $0.0027^{\dagger}$ & 70.3\%\\
Strain heterogeneity & -0.374 & [-0.585,-0.203] & $0.0286$ & 5.3\%\\
Relative excess length & -0.308 & [-0.511,-0.132] & $0.0704$ & 2.1\%\\
Coordination anomalies per length & 0.246 & [0.068,0.468] & $0.2388$ & 1.7\%\\
Grain-area imbalance & -0.350 & [-0.561,-0.158] & $0.0704$ & 3.8\%\\
Misorientation & 0.120 & [-0.053,0.256] & $0.2470$ & 0.4\%\\
\bottomrule\end{tabular}

\caption{Exploratory associations after size adjustment. A dagger marks the Monte Carlo resolution floor: zero exceedances among 10000 permutations give $p_{\rm MC}=1/10001$ and minimum Holm-adjusted output $27/10001\simeq0.0027$. These are resolution-limited values, not exact tail probabilities or rigorous upper bounds on the underlying permutation probability.}\label{tab:effective-associations}
\end{table*}

Power fits depend strongly on reference: the prescribed-crystal exponent is about 1.37 and the chemical-potential-matched exponent about 0.87. Neither identifies an asymptotic law.

Changing density convention shifts the excess by up to $7.593\times10^{-4}$; changing bulk exclusion depth shifts it by up to $5.152\times10^{-5}$. Chemical-potential matching reduces the mismatch below $8.756\times10^{-10}$. These are deterministic comparison sensitivities.

For the twelve-state matched group, omitting one size gives slopes from 0.11151 to 0.11320. Two of three mean residual summaries exclude zero after Bonferroni adjustment over the three omitted sizes within the half-trace outcome. The proportional-fit percentile interval $[0.11222,0.11321]$ is conditional on the cohort, reference and three sizes. With three to five states per size, it has no established population coverage and excludes systematic size and reference effects. The power-fit conditional percentiles are $[1.022,1.046]$, consistent with the departures from strict proportionality. Native-density and deeper-interior references give slopes 0.11302 and 0.11278; alternate overlapping cohorts give 0.11154 and 0.11071.

\section{Matching grain interiors}\label{app:matching}

The local Green strain is $E=(F^{\mathsf T}F-I)/2$, where $I$ is the identity matrix; $E_g$ denotes its area-weighted mean over ordered interior sites in grain $g$. The local $F$ is fitted to the six neighbor displacements relative to the direct seed basis of Appendix~\ref{sec:preparations}. A superscript $\mathrm{ref}$ on a strain component denotes the corresponding ideal-template value. Mutual matching requires every pair to satisfy absolute tolerances of 0.004 for Green-strain diagonals, 0.002 for shear, 0.002 for each grain density, $10^{-3}$ for chemical potential, $0.2^\circ$ for misorientation and individual grain orientation, 0.03 for grain-area fraction, and $10^{-4}$ for bulk-tensor components. Relative cell-area differences are at most 1\%; the mean interior-density target tolerance is $2\times10^{-4}$. Cohorts are ranked by number of distinct sizes, then number of independent states, with preparation identifiers breaking ties. The defective response does not enter selection.

The ordered-template construction, density normalization and twelve-state cohort are described in Sec.~\ref{sec:controlled}. Density adjustments are followed by renewed minimization. Direct checks of the physical fields verify stationarity and the identity $\mathcal E[\psi]=n_0\mathcal E_{\Lambda_\phi}[\phi]$, where $\av{|\phi|^2}=1$ and $\mathcal E_{\Lambda_\phi}$ is Eq.~\eqref{eq:energy} with kernel amplitude $n_0U_0$. The twenty-eight converged outcomes comprise four preparations at the smallest box and eight at each larger box. Every pair in a cohort must meet all tolerances; a chain of pairwise matches is insufficient. None of the four $m=1$ states at $L=135.019$ enters this cohort: their relaxed normal strains do not overlap its simultaneous interior-condition window; one also fails the orientation criterion. These states are stationary outcomes outside the matched cohort. Halving all pairwise tolerances leaves a cohort spanning only two sizes.

\begin{table*}[!t]\centering\small\begin{tabular}{rrrrrr}\toprule $m$ & Preparation & $L$ & $n_0$ & $n_{\rm bulk}$ & Matched cohort\\\midrule
1 & 0 & 135.019 & 0.980880 & 1.000000 & no\\
1 & 1 & 135.019 & 0.981785 & 0.999999 & no\\
1 & 2 & 135.019 & 0.980936 & 0.999999 & no\\
1 & 3 & 135.019 & 0.980890 & 0.999998 & no\\
2 & 0 & 270.038 & 0.990654 & 1.000000 & yes\\
2 & 1 & 270.038 & 0.990621 & 1.000000 & yes\\
2 & 2 & 270.038 & 0.990662 & 1.000000 & no\\
2 & 3 & 270.038 & 0.990618 & 1.000000 & yes\\
2 & 4 & 270.038 & 0.990650 & 1.000000 & no\\
2 & 5 & 270.038 & 0.990600 & 1.000000 & no\\
2 & 6 & 270.038 & 0.990629 & 1.000000 & no\\
2 & 7 & 270.038 & 0.990626 & 1.000000 & yes\\
3 & 0 & 405.056 & 0.993790 & 1.000000 & yes\\
3 & 1 & 405.056 & 0.993772 & 1.000000 & no\\
3 & 2 & 405.056 & 0.993751 & 1.000000 & no\\
3 & 3 & 405.056 & 0.993769 & 1.000000 & no\\
3 & 4 & 405.056 & 0.993774 & 1.000000 & yes\\
3 & 5 & 405.056 & 0.993793 & 1.000000 & yes\\
3 & 6 & 405.056 & 0.993804 & 1.000000 & no\\
3 & 7 & 405.056 & 0.993847 & 1.000000 & no\\
4 & 0 & 540.075 & 0.995328 & 1.000000 & yes\\
4 & 1 & 540.075 & 0.995327 & 1.000000 & yes\\
4 & 2 & 540.075 & 0.995350 & 1.000000 & yes\\
4 & 3 & 540.075 & 0.995346 & 1.000000 & yes\\
4 & 4 & 540.075 & 0.995358 & 1.000000 & no\\
4 & 5 & 540.075 & 0.995521 & 1.000158 & no\\
4 & 6 & 540.075 & 0.995366 & 1.000000 & no\\
4 & 7 & 540.075 & 0.995331 & 1.000000 & yes\\
\bottomrule\end{tabular}

\caption{Ordered-template ensemble and interior-matched subset. $m$ is the box multiple, $L=mL_0$, and $n_{\rm bulk}$ the measured grain-interior density. All converged states are listed; a mismatch against the ideal template is distinct from mutual comparability between relaxed preparations.}\label{tab:matched-pilot}
\end{table*}

Within this group, maximum differences are $0.00382$ in a Green-strain diagonal, $2.07\times10^{-4}$ in physical grain density, $8.88\times10^{-4}$ in chemical potential and $4.09\times10^{-5}$ in a homogeneous bulk-tensor component. Conditions are approximately matched within these tolerances, not identical.

\begin{figure*}[!t]\centering\includegraphics[width=\textwidth]{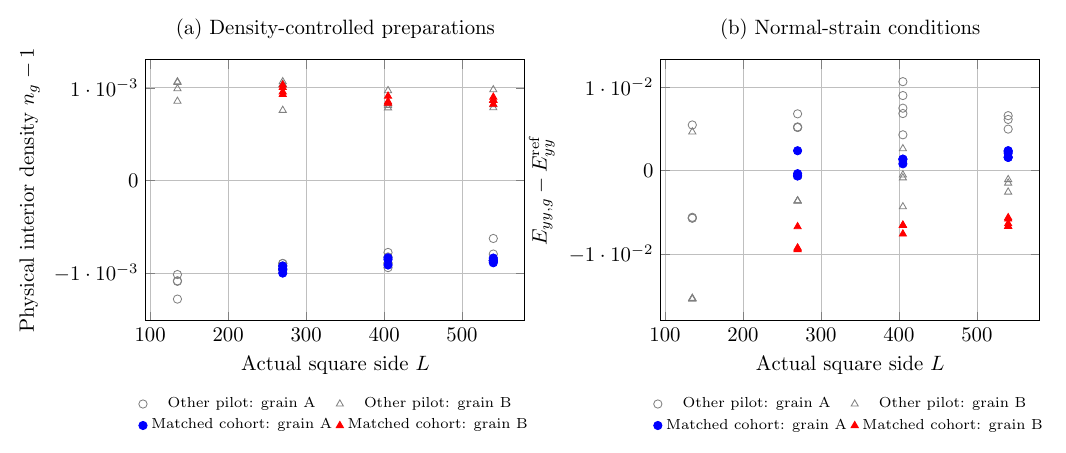}
\caption{Interior densities and normal strains in the ordered-template family. Colored symbols identify the twelve matched preparations; gray symbols show the remaining states.}\label{fig:interior-matching}
\end{figure*}

\section{Larger-box and fixed-area tests}\label{app:validation}
\subsection{Larger-box comparison}
The coefficient $\Gamma_K=0.11272$ of Sec.~\ref{sec:controlled} is compared, without refitting, with eight independent preparations at $L=675.094$. All fields relax freely, and six reach $R\leq10^{-9}$. Two of these satisfy every pairwise tolerance with the full twelve-state group; three satisfy them with respect to the $L=540.075$ states and are used here. Define the relative prediction error $\varepsilon_{\rm pred}=\langle\Delta\overline K\rangle_{\rm prep}/[\Gamma_K\langle\nu\rangle_{\rm prep}]-1$. Its half-trace mean is $-2.45\%$, with conditional bootstrap percentiles $[-2.87,-1.99]\%$ that include resampling of the fitted coefficient and of the preparations. The exclusion of zero indicates systematic overprediction, consistent with the decrease of $\Gamma_{K,\rm eff}$ with $L$. The transverse error is $-4.48\%$, with percentiles $[-4.88,-4.03]\%$. Three states compared with a single size do not support a population-level validation.

\subsection{Fixed-area test}\label{app:paired-length}
Eight independent blocks at fixed $L=405.056276$ and $N_g=864$ each contain one state with two and one with four initial interfaces. Grain translations and roughness and noise phases are shared within a block and independent between blocks. The physical interaction is that of Sec.~\ref{sec:controlled}, density feedback targets grain-interior density one, and both fields relax freely without pinning. All sixteen final states satisfy $R\leq10^{-9}$; direct checks of the physical fields verify mass, stationarity and the density-rescaling identity. All states keep their two or four interfaces.

\begin{figure*}[!t]\centering\includegraphics[width=\textwidth]{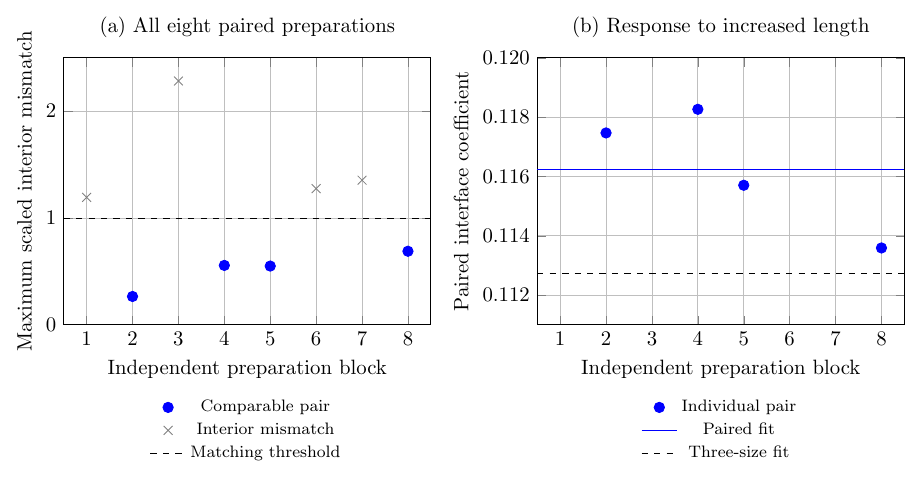}
\caption{Fixed-area interface-length test. (a) Maximum interior mismatch divided by its allowed tolerance for each block; the line at one is the acceptance threshold. (b) Individual paired slopes for the four accepted blocks, with the paired fit and the three-size coefficient of Sec.~\ref{sec:controlled} as horizontal lines. Individual points carry no resampling error bars. Selection depends on bulk conditions and geometry, not on the response.}\label{fig:paired-length}
\end{figure*}

A pair is accepted if its two states satisfy the interior-condition tolerances of Appendix~\ref{app:matching} and their measured length ratio is at least 1.25; the defective response is not used. Four of the eight pairs are accepted; the other four exceed the strain-diagonal and chemical-potential tolerances. Figure~\ref{fig:paired-length}(a) shows all eight blocks. In the accepted pairs, measured length ratios range from 1.99435 to 2.00136.

\begin{table*}[!t]\centering\small\begin{tabular}{rrrr}\toprule Block & $\delta\nu$ & $\delta K$ & $\delta K/\delta\nu$\\\midrule
2 & 0.005069 & 0.0005954 & 0.1175\\
4 & 0.005084 & 0.0006012 & 0.1183\\
5 & 0.005084 & 0.0005883 & 0.1157\\
8 & 0.005098 & 0.0005790 & 0.1136\\
\bottomrule\end{tabular}

\caption{Fixed-area pairs satisfying the within-pair criteria. Blocks are numbered 1--8. $\delta K$ is the increment in bulk-subtracted half-trace physical stiffness. These four blocks do not all share the same interior conditions.}\label{tab:paired-length}
\end{table*}

All four half-trace increments and both directional increments are positive. Their proportional slope is 0.1162, with mean individual slope 0.1163 and preparation standard deviation 0.002076. Resampling the four independent blocks gives conditional percentile endpoints $[0.11455,0.11787]$; this small-sample diagnostic has unestablished population coverage and excludes reference uncertainty. The one-sided exact sign-test probability for four positive independent increments is $2^{-4}=0.0625$, the smallest attainable value with four pairs, so the slope dispersion supplies the descriptive comparison. The measured $\delta\nu$ values span only 0.0050691--0.0050976, so this test probes a two-level length change rather than an exponent or an intercept.

Requiring mutual comparability between every state of different blocks leaves blocks 2 and 5, whose slope is 0.11658. With two blocks this comparison is descriptive only; the within-pair and cross-block comparisons answer distinct questions.

\subsection{Domain of the fixed-area comparison}\label{app:three-level}
At $L=675.09$, no paired triplet with two, four and six interfaces simultaneously satisfies stationarity and the interior-matching criteria. The fixed-area evidence of Appendix~\ref{app:paired-length} therefore resolves a two-level length change without constraining curvature with interface density in the larger box.

\section{Absolute perfect-crystal baseline diagnostics}\label{app:baseline}\label{sec:baseline}
Table~1 of Ref.~\cite{Sepulveda2010} reports perfect-crystal eigenvalues $f_1=f_2=0.051$ at $\Lambda=128.68$, $a=8$ and box side $L=256$. Our specified reference at those stated physical parameters has $\fbar=0.00912955$, approximately 5.59 times smaller. The parameter conventions agree; this absolute baseline is not quantitatively reproduced.

Two controlled continuations are computed at fixed $L=256$, $a=8$ and $\Lambda=128.68$. Refining the commensurate reference from $h=1$ through $0.5$ to $0.25$ gives $\fbar\simeq0.00912952$. Sampling the monocrystal seed of Eq.~(21) of Ref.~\cite{Sepulveda2010} with the stated wave number $q_0=0.65$ and amplitude 0.1 instead yields a defect-free 598-site branch with $\fbar\simeq0.00777337$ after refinement. An equilateral primitive cell at the same wave number gives $0.00780261$. Thus the low response persists under refinement and this alternative seed, rather than recovering 0.051.

At $h=2$ the commensurate reference gives $\fbar=0.0208162$ and a strongly altered directional tensor. This demonstrates a possible numerical mechanism for inflated responses, but this tested coarse mesh does not reproduce the published isotropic 0.051. The caption of Table~1 of Ref.~\cite{Sepulveda2010} states $256\times256$ \emph{units}; it does not specify the number of Fourier grid points or spacing for those runs. Moreover, the rounded $q_0=0.65$ seed is not exactly commensurate with that square torus. Our sampled-seed calculation is consequently a diagnostic, not a reconstruction of the published discretization or microscopic state.

A supplementary relaxation check in the equilateral $q_0=0.65$ cell starts from a triangular seed of amplitude $\epsilon=0.1$ and evaluates saved prefixes of a single deterministic minimization. After six iterations the state has residual $R=0.0597$, as defined in Eq.~\eqref{eq:residual}, an energy 1.6\% above the stationary value, and a phase response of $0.05454$; continuing the same trajectory to $R\leq10^{-11}$ lowers the response to $0.007807$. The response is evaluated independently from the density using periodic conforming finite elements; doubling the density resolution leaves these values unchanged at the reported precision. Incomplete relaxation can therefore produce responses close to the published value, offering a possible numerical explanation of the difference. The intermediate state is not an equilibrium, and this check does not identify the origin of the published value.

Blakie~\cite{Blakie2024}, in the subsection entitled ``Superfluid fraction results'', also reported differences from the homogeneous soft-core results of Ref.~\cite{Sepulveda2010} and discussed commensuration strain and sharply varying auxiliary phase fields as numerical sensitivities. Ref.~\cite{Sepulveda2010} does not specify the branch and discretization of that calculation; the present checks establish numerical sensitivities without identifying the origin of the published value.

The seven square-domain states in Table~\ref{tab:baseline-diagnostics} use the spectral solver of Appendix~\ref{app:numerics} and the same physical interaction. The commensurate branch is initialized from the converged $L=256$ reference, with downsampling for coarse grids and spectral continuation for refinement. The second chain samples the monocrystal seed of Eq.~(21) of Ref.~\cite{Sepulveda2010} with the rounded wave number $q_0=0.65$, normalizes its mean density to one, and then refines that relaxed state. Direct checks of the fields verify norm and $R\leq10^{-9}$ for all seven states. Each chain is a numerical continuation, not a statistical ensemble.

\begin{table*}[!t]\centering\small\begin{tabular}{lrrrrr}\toprule Initialization & $N_g$ & $h$ & $\fbar$ & $N_s$ & $N_{\ne6}$\\\midrule
Commensurate & 128 & 2 & 0.02081620 & 672 & 0\\
Commensurate & 256 & 1 & 0.00912934 & 672 & 0\\
Commensurate & 512 & 0.5 & 0.00912952 & 672 & 0\\
Commensurate & 1024 & 0.25 & 0.00912952 & 672 & 0\\
Eq.~(21), $q_0=0.65$ & 256 & 1 & 0.00777339 & 598 & 0\\
Eq.~(21), $q_0=0.65$ & 512 & 0.5 & 0.00777337 & 598 & 0\\
Eq.~(21), $q_0=0.65$ & 1024 & 0.25 & 0.00777337 & 598 & 0\\
\bottomrule\end{tabular}

\caption{Baseline diagnostics on the $L=256$ square torus. Both chains retain sixfold coordination but have different site counts. These calculations do not identify the mesh used in Ref.~\cite{Sepulveda2010}.}\label{tab:baseline-diagnostics}
\end{table*}

As a separate geometric check, an exactly equilateral primitive cell with reciprocal wave number 0.65 is solved with $16^2$, $32^2$, $64^2$ and $128^2$ spectral points. Its response converges to $0.00780261$. This checks a different admissible crystal cell and the primitive-cell response calculation; it does not select the branch of that Table~1. The coarse square-grid enhancement, the converged low-response branches and the equilateral calculation constrain explanations but do not establish that the published calculation lacked resolution.

\section{Reduced transport model and coupling dependence}\label{app:transport-model}
\subsection{Link extraction and state-level comparisons}
Link densities are evaluated by periodic cubic interpolation of the positive GP amplitude followed by squaring. Density samples resolve the minimum along each Delaunay bond; the graph phases are minimized with one gauge fixed. The stationary graph residual is recomputed after convergence. A periodic square graph gives the exact uniform response, and node relabeling, rotations and uniform rescaling preserve the corresponding tensor identities.

The interior relation in Eq.~\eqref{eq:link-law} is fitted with equal total weight per state. The periodically repeated reference at $L=256$ is omitted from this fit because it replicates the $L=128$ crystal. Leaving out a whole state gives similar logarithmic prediction errors for linear and squared-distance fits, so the available interval does not distinguish their asymptotic forms. Individual bonds are not treated as independent preparations.

For any tensor $T$, define $\|T\|_{\max}=\max_{i,j}|T_{ij}|$. GP denotes the continuous Gross--Pitaevskii response, and net the reduced network response. The tensor error is measured by $\|K^{\rm net}-K^{\rm GP}\|_{\max}/\|K^{\rm GP}\|_{\max}$, not by dividing each component by itself. It is at most 5.1\% for the eight illustrative states at $\Lambda=128.68$. The 6.0\% entry at the same coupling in Table~\ref{tab:coupling-transfer} refers to the separate paired-branch states and their reference. Neither value implies the same relative accuracy for small off-diagonal components or bulk-subtracted excesses. Table~\ref{tab:network-excess} illustrates the amplification under subtraction. The reduced model captures the scale of the parallel enhancement but can reverse a small transverse excess or a near-zero half-trace excess.
\begin{table*}[!t]\centering\small\begin{tabular}{lrrrr}\toprule
State & $\Delta K_{xx}^{\rm GP}$ & $\Delta K_{xx}^{\rm net}$ & $\Delta K_{yy}^{\rm GP}$ & $\Delta K_{yy}^{\rm net}$\\
 & \multicolumn{4}{c}{$(10^{-3})$}\\\midrule
$30^\circ$, $L=128$ & 2.454 & 1.919 & 0.07536 & -0.1786\\
$30^\circ$, $L=256$ & 1.706 & 1.611 & -0.2037 & -0.3425\\
$15.75^\circ$, $L=256$ & 0.2166 & -0.1126 & -0.2115 & -0.3924\\
\bottomrule\end{tabular}

\caption{Network and GP excesses under the same prescribed comparators. The $30^\circ$ states use their prescribed single-crystal reference; the low-angle state uses the arithmetic mean of the two 672-site references. These comparators differ from the recommended matched composite and isolate errors introduced by the reduced model.}\label{tab:network-excess}
\end{table*}

A constant-width channel alternative uses $J_b\propto[\int_b\dd\sigma/\rho(\sigma)]^{-1}$, with $\sigma$ the arc length along the bond, with a separate single-crystal calibration. Its maximum normalized tensor error is 9.4\% on these states, compared with 5.1\% for the minimum-density model. The comparison supports the empirical usefulness of link minima in this regime without deriving a universal bond conductance. Replacing the coarse density by the refined GP field on the fixed graphs changes representative network tensors by at most 0.22\%; direct GP stiffness changes are below $6\times10^{-10}$. The extraction error is smaller than the discrepancy between the reduced and continuous descriptions.

\subsection{Fixed-network interventions}
Compressed boundary bonds have midpoint distance less than $a$ from a non-sixfold site, length below the minimum interior length of the same state, and coupling above its interior median. Their couplings are capped at that median with every other coupling and all node positions fixed. The decrease of the effective tensor obeys graph-response monotonicity; this decrease alone does not identify a preferential transport direction.

For the directional comparison, equal numbers of boundary and interior bonds are selected independently and their couplings halved. Thirty-two subsets of each class define the mean influence shown in Fig.~\ref{fig:link-network}c. Boundary-to-interior influence ratios are 1.10--1.47 for parallel transport and 0.51--0.89 for transverse transport. These are conditional graph interventions, not additional stationary GP states or independent preparation samples.

\subsection{Interaction and deformation comparisons}
Two independently prepared $L=128$ bicrystals define paired branches at the three couplings in Table~\ref{tab:coupling-transfer}; the mean density remains one and $a=8$. All six bicrystals satisfy the same two-boundary structural criteria. Grain deformations, densities and partitions are remeasured for each stationary state, and each homogeneous comparator is matched to its chemical potential. The coupling enters the physical cell solve as $\Lambda_0n_g$. The paired branches are not counted as independent preparations at different couplings. The prescribed reference retains the same commensurate cell at all three couplings. The equilateral benchmark of Appendix~\ref{app:external-benchmark} minimizes the cell energy and defines a separate crystalline branch.
\begin{table*}[!t]\centering\small\begin{tabular}{rrrrrr}\toprule
$\Lambda$ & $\overline K_{\rm ref}$ & $\Delta\overline K_1$ & $\Delta\overline K_2$ & Dilation response & Network error\\\midrule
80 & 0.08030 & 0.01135 & 0.01240 & $-7.3\%$ & 15.1\%\\
100 & 0.03042 & 0.008369 & 0.007867 & $-6.1\%$ & 6.9\%\\
128.68 & 0.009130 & 0.002555 & 0.002422 & $-4.7\%$ & 6.0\%\\
\bottomrule\end{tabular}

\caption{Interaction dependence of the GP response and reduced model. $\overline K_{\rm ref}$ is the prescribed commensurate reference at unit density. The two excesses use their own strain- and chemical-potential-matched composites. Dilation response is the relative change of the prescribed reference under a 1\% increase of both cell vectors at fixed density. Network error is the largest normalized tensor error among the two bicrystals and the reference, retaining the calibration at $\Lambda=128.68$.}\label{tab:coupling-transfer}
\end{table*}

Halving the spatial spacing from 0.5 to 0.25 in one paired state at each coupling preserves the positive excess. The largest change in the complete half-trace excess is $1.1\times10^{-6}$, less than 0.01\% of its value. Primitive-cell deformation responses are unchanged at the quoted precision between $32^2$ and $64^2$ points. A finite-dimensional amplitude-curvature check in twenty smooth collective directions, with norm and global translations projected out, gives positive curvature for representative bicrystals at all three couplings and for the prescribed reference. This tests those collective directions and is not a full Bogoliubov stability analysis.

\clearpage
\nocite{*}
\bibliographystyle{apsrev4-2}
\bibliography{aps-controls,references}

@CONTROL{prbBibliographyControl,
 title = {0}}

@article{Sepulveda2010,
 author={Sep{\'u}lveda, N. and Josserand, C. and Rica, S.},
 title={Superfluid density in a two-dimensional model of supersolid},
 journal={The European Physical Journal B}, volume={78}, pages={439--447}, year={2010},
 doi={10.1140/epjb/e2010-10176-y}}

@article{Leggett1970,
 author={Leggett, A. J.}, title={Can a Solid Be ``Superfluid''?},
 journal={Physical Review Letters}, volume={25}, pages={1543}, year={1970},
 doi={10.1103/PhysRevLett.25.1543}}

@article{Pollet2007,
 author={Pollet, L. and Boninsegni, M. and Kuklov, A. B. and Prokof'ev, N. V. and Svistunov, B. V. and Troyer, M.},
 title={Superfluidity of Grain Boundaries in Solid Helium-4},
 journal={Physical Review Letters}, volume={98}, pages={135301}, year={2007},
 doi={10.1103/PhysRevLett.98.135301}}

@article{Cinti2014,
 author={Cinti, F. and Macr{\`i}, T. and Lechner, W. and Pupillo, G. and Pohl, T.},
 title={Defect-induced supersolidity with soft-core bosons},
 journal={Nature Communications}, volume={5}, pages={3235}, year={2014},
 doi={10.1038/ncomms4235}}

@article{Zhang2022,
 author={Zhang, Jin and Zhang, Chao and Yang, Jin and Capogrosso-Sansone, Barbara},
 title={Supersolid phases of lattice dipoles tilted in three dimensions},
 journal={Physical Review A}, volume={105}, pages={063302}, year={2022},
 doi={10.1103/PhysRevA.105.063302}}

@article{Norcia2021,
 author={Norcia, Matthew A. and Politi, Claudia and Klaus, Lauritz and Poli, Elena and Sohmen, Maximilian and Mark, Manfred J. and Bisset, Russell N. and Santos, Luis and Ferlaino, Francesca},
 title={Two-dimensional supersolidity in a dipolar quantum gas},
 journal={Nature}, volume={596}, pages={357--361}, year={2021},
 doi={10.1038/s41586-021-03725-7}}

@article{Tao2023,
 author={Tao, J. and Zhao, M. and Spielman, I. B.},
 title={Observation of Anisotropic Superfluid Density in an Artificial Crystal},
 journal={Physical Review Letters}, volume={131}, pages={163401}, year={2023},
 doi={10.1103/PhysRevLett.131.163401}}

@article{LiuNocedal1989,
 author = {Liu, Dong C. and Nocedal, Jorge},
 title = {On the limited memory {BFGS} method for large scale optimization},
 journal = {Mathematical Programming},
 volume = {45},
 pages = {503--528},
 year = {1989},
 doi = {10.1007/BF01589116}
}

@book{Briggs2000,
 author = {Briggs, William L. and Henson, Van Emden and McCormick, Steve F.},
 title = {A Multigrid Tutorial},
 edition = {2},
 publisher = {Society for Industrial and Applied Mathematics},
 address = {Philadelphia},
 year = {2000},
 doi = {10.1137/1.9780898719505}
}

@article{Pomeau1994,
 author={Pomeau, Yves and Rica, Sergio}, title={Dynamics of a model of supersolid},
 journal={Physical Review Letters}, volume={72}, pages={2426--2429}, year={1994},
 doi={10.1103/PhysRevLett.72.2426}}

@article{Josserand2007,
 author={Josserand, Christophe and Pomeau, Yves and Rica, Sergio},
 title={Coexistence of Ordinary Elasticity and Superfluidity in a Model of a Defect-Free Supersolid},
 journal={Physical Review Letters}, volume={98}, pages={195301}, year={2007},
 doi={10.1103/PhysRevLett.98.195301}}

@article{Sepulveda2008,
 author={Sep{\'u}lveda, N{\'e}stor and Josserand, Christophe and Rica, Sergio},
 title={Nonclassical rotational inertia fraction in a one-dimensional model of a supersolid},
 journal={Physical Review B}, volume={77}, pages={054513}, year={2008},
 doi={10.1103/PhysRevB.77.054513}}

@article{Blakie2024,
 author={Blakie, P. Blair}, title={Superfluid fraction tensor of a two-dimensional supersolid},
 journal={Journal of Physics B: Atomic, Molecular and Optical Physics},
 volume={57}, pages={115301}, year={2024}, doi={10.1088/1361-6455/ad41c1},
 eprint={2308.14001}, archivePrefix={arXiv}}

@article{Saslow1976,
 author={Saslow, W. M.}, title={Superfluidity of periodic solids},
 journal={Physical Review Letters}, volume={36}, pages={1151--1154}, year={1976},
 doi={10.1103/PhysRevLett.36.1151}}

@article{Leggett1998,
 author={Leggett, A. J.}, title={On the Superfluid Fraction of an Arbitrary Many-Body System at {$T=0$}},
 journal={Journal of Statistical Physics}, volume={93}, pages={927--941}, year={1998},
 doi={10.1023/B:JOSS.0000033170.38619.6c}}

@article{Chauveau2023,
 author={Chauveau, G. and Maury, C. and Rabec, F. and Heintze, C. and Brochier, G. and Nascimbene, S. and Dalibard, J. and Beugnon, J. and Roccuzzo, S. M. and Stringari, S.},
 title={Superfluid Fraction in an Interacting Spatially Modulated {Bose-Einstein} Condensate},
 journal={Physical Review Letters}, volume={130}, pages={226003}, year={2023},
 doi={10.1103/PhysRevLett.130.226003}}

@article{Biagioni2024,
 author={Biagioni, G. and Antolini, N. and Donelli, B. and Pezz{\`e}, L. and Smerzi, A. and Fattori, M. and Fioretti, A. and Gabbanini, C. and Inguscio, M. and Tanzi, L. and Modugno, G.},
 title={Measurement of the superfluid fraction of a supersolid by {Josephson} effect},
 journal={Nature}, volume={629}, pages={773--777}, year={2024},
 doi={10.1038/s41586-024-07361-9}}

@article{Henkel2010,
 author={Henkel, N. and Nath, R. and Pohl, T.},
 title={Three-Dimensional Roton Excitations and Supersolid Formation in {Rydberg}-Excited {Bose-Einstein} Condensates},
 journal={Physical Review Letters}, volume={104}, pages={195302}, year={2010},
 doi={10.1103/PhysRevLett.104.195302}}

@article{Cinti2010,
 author={Cinti, F. and Jain, P. and Boninsegni, M. and Micheli, A. and Zoller, P. and Pupillo, G.},
 title={Supersolid Droplet Crystal in a Dipole-Blockaded Gas},
 journal={Physical Review Letters}, volume={105}, pages={135301}, year={2010},
 doi={10.1103/PhysRevLett.105.135301}}

@article{Macri2013,
 author={Macr{\`i}, T. and Maucher, F. and Cinti, F. and Pohl, T.},
 title={Elementary excitations of ultracold soft-core bosons across the superfluid-supersolid phase transition},
 journal={Physical Review A}, volume={87}, pages={061602(R)}, year={2013},
 doi={10.1103/PhysRevA.87.061602}}

@article{Boninsegni2012,
 author={Boninsegni, Massimo and Prokof'ev, Nikolay V.},
 title={Colloquium: Supersolids: What and where are they?},
 journal={Reviews of Modern Physics}, volume={84}, pages={759--776}, year={2012},
 doi={10.1103/RevModPhys.84.759}}

@article{Gibbs1878,
 author={Gibbs, J. Willard},
 title={On the Equilibrium of Heterogeneous Substances},
 journal={Transactions of the Connecticut Academy of Arts and Sciences},
 volume={3}, pages={108--248, 343--524}, year={1876--1878}}

@incollection{Cahn1979,
 author={Cahn, John W.}, title={Thermodynamics of Solid and Fluid Surfaces},
 booktitle={Interfacial Segregation}, editor={Johnson, W. C. and Blakely, J. M.},
 publisher={American Society for Metals}, address={Metals Park, Ohio},
 pages={3--23}, year={1979}}

@article{Rakic2024,
 author={Rakic, Milan and Ho, Andrew F. and Lee, Derek K. K.},
 title={Elastic properties and thermodynamic anomalies of supersolids},
 journal={Physical Review Research}, volume={6}, pages={043040}, year={2024},
 doi={10.1103/PhysRevResearch.6.043040}, eprint={2403.13727}, archivePrefix={arXiv}}

@article{Peotta2025,
 author={Peotta, Sebastiano and Spada, Gabriele and Giorgini, Stefano and Pilati, Sebastiano and Recati, Alessio},
 title={Supersolid phase in two-dimensional soft-core bosons at finite temperature},
 journal={Physical Review B}, volume={112}, pages={184506}, year={2025},
 doi={10.1103/syns-jm6w}, eprint={2504.03482}, archivePrefix={arXiv}}
\end{document}